\documentclass[12pt]{article}

\usepackage{braket}
\usepackage[margin=0.7in]{geometry}
\usepackage{authblk}
\usepackage{amsmath}
\usepackage{amssymb}
\usepackage{graphicx}
\usepackage[font=footnotesize,labelfont=bf, margin = 0.7in]{caption}
\usepackage{lineno}
\usepackage{titlesec}
\titleformat*{\subsection}{\normalfont\fontfamily{phv}\fontsize{14}{17}\normalfont}
\newtheorem{theorem}{Theorem}
\newtheorem{lemma}[theorem]{Lemma}

\begin{document}

\title{Quantum Physics in Connected Worlds}

\author[1,2, *]{Joseph Tindall}
\author[2]{Amy Searle}
\author[3]{Abdulla Alhajri}
\author[4,5,2]{Dieter Jaksch}
\affil[1]{Center for Computational Quantum Physics, Flatiron Institute, 162 5th Avenue, New York, NY 10010.}
\affil[2]{Clarendon Laboratory, University of Oxford, Parks Road, Oxford OX1 3PU, United Kingdom}
\affil[3]{Technology Innovation Institute, Masdar City 9639,
Abu Dhabi, United Arab Emirates}
\affil[4]{The Hamburg Centre for Ultrafast Imaging, Universität Hamburg, Luruper Chaussee 149, 22761 Hamburg, Germany}
\affil[5]{Institut für Laserphysik, Universität Hamburg, Luruper Chaussee 149, 22761 Hamburg, Germany}
\affil[*]{Corresponding author. Email: jtindall@flatironinstitute.org}

\maketitle

\abstract{Theoretical research into many-body quantum systems has mostly focused on regular structures which have a small, simple unit cell and where a vanishingly small number of pairs of the constituents directly interact. Motivated by advances in control over the pairwise interactions in many-body simulators, we determine the fate of spin systems on more general, arbitrary graphs. Placing the minimum possible constraints on the underlying graph, we prove how, with certainty in the thermodynamic limit, such systems behave like a single collective spin. We thus understand the emergence of complex many-body physics as dependent on `exceptional', geometrically constrained structures such as the low-dimensional, regular ones found in nature. Within the space of dense graphs we identify hitherto unknown exceptions via their inhomogeneity and observe how complexity is heralded in these systems by entanglement and highly non-uniform correlation functions. Our work paves the way for the discovery and exploitation of a whole class of geometries which can host uniquely complex phases of matter.}
\newline 
\newline

\section*{Introduction}
{\fontsize{11pt}{13.2pt}\selectfont Research into many-body quantum physics has predominantly involved setups with local interactions and a high degree of spatial symmetry. Such a focus is natural, with the short-range, homogeneous nature of the resulting Hamiltonian being a reasonable reflection of reality in naturally occurring materials, and also beneficial, since such features can be exploited in order to render the Hamiltonian soluble with computational methods. 

Recent experimental advances, however, have made it clear that many-body quantum physics need not be limited to such geometries. In Rydberg simulators \cite{RydbergTweezers1, RydbergTweezers2, RydbergTweezers3, RydbergTweezers5, RydbergTweezers6}, for example, free placement of the individual atoms is now possible using optical tweezers. Moreover, in a range of other platforms --- which include atoms trapped in cavities \cite{InteractionControl5} or photonic waveguides \cite{InteractionControl1}, Moir\'e Heterostructures \cite{InteractionControl6}, trapped ions \cite{InteractionControl2} and superconducting circuits \cite{InteractionControl7} --- experimentalists are demonstrating increasing control over the pairwise interactions and geometries in the Hamiltonians that they can realise. For instance, proposals now exist to use trapped ion arrays to engineer many-body spin Hamiltonians defined over arbitrary graphs \cite{InteractionControlArbGraph, InteractionControl3} whilst a recent experiment successfully probed the out-of-equilibrium behaviour of a spin model with all-to-all interactions \cite{AlltoAllLMGExperiment}.
\par In general, the limitations on the geometries which can be realised are continually being lowered and open up the tantalising possibility of exploring and utilising many-body quantum physics on a wide range of complex graph structures, including those which are well-established in the social \cite{SocialScienceGraphTheory2} and biological \cite{BiologyGraphTheory1, NeuroscienceGraphTheory1} sciences. 

\begin{figure}[!t]
    \centering
    \includegraphics[width =0.8\textwidth]{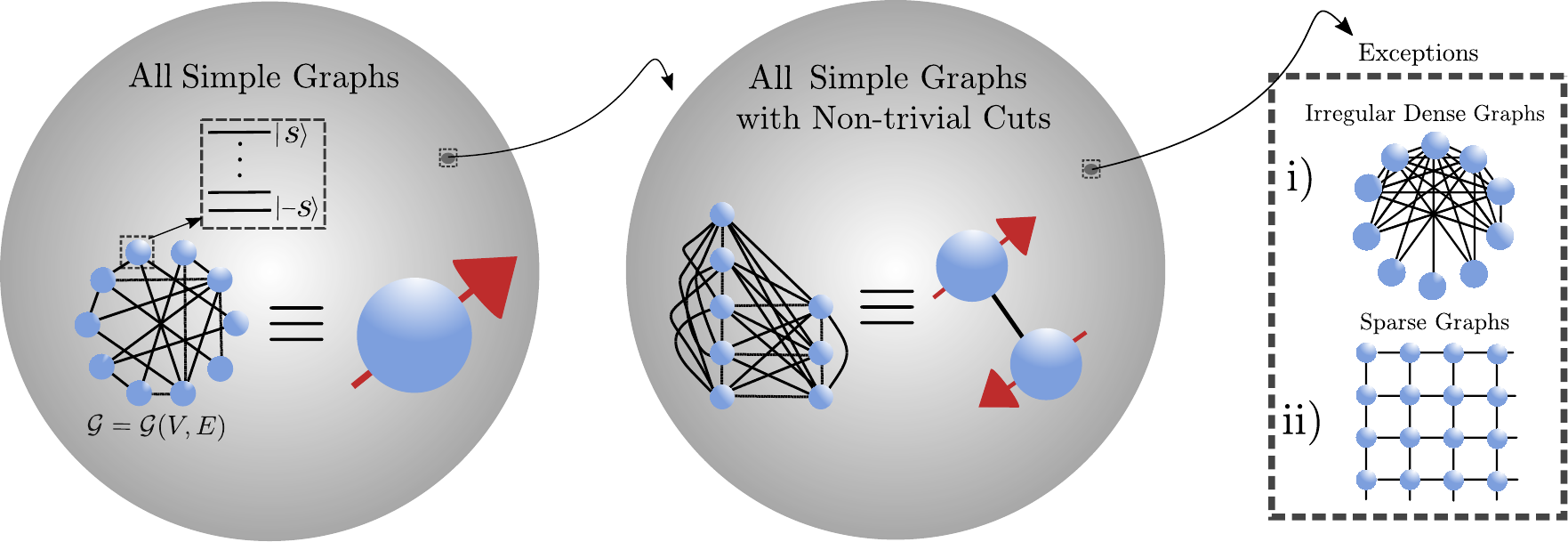}
    \caption{In this work we consider a many-body spin $s$ Hamiltonian --- see Eq. (\ref{Eq:GeneralHamiltonian}) --- defined over a simple graph $\mathcal{G}(V,E)$ where the vertices represent the spins and the edges the pairs of spins upon which the two-body terms act. We prove for a graph chosen uniformly at random from all simple graphs that, as the graph size $L$ increases, the equilibrium properties of the system become increasingly like that of a single collective spin and any many-body effects vanish as $L \rightarrow \infty$. In order for this not to be true the graph must possess a non-trivial cut and we prove that even for such a graph, chosen at random amongst all those with a non-trivial cut, the system can effectively be reduced to that of two interacting collective spins. 
    The emergence of complex, non-collective physics is thus dependent on more structured, `exceptional' graphs which exist in a vanishingly small subspace of the space of all simple graphs. These include the well-known sparse, regular structures that arise in nature and a new class of graphs we identify here: irregular dense structures.}
    \label{Fig:F1}
\end{figure}

\par Despite this experimental progress, from a theoretical perspective there is little understanding of the physics of many-body Hamiltonians when hosted on structures which are not either all-to-all setups \cite{SYKSpinGlass1} or sparsely connected, low-dimensional lattices. The last decade has seen significant interest in low-dimensional lattices with long-range interactions \cite{PowerLaw1, PowerLaw2, EqmPowerLaw3, OutofEqmPowerLaw1, OutofEqmPowerLaw5}, yet despite the increased connectivity the underlying system is still translationally invariant. The fate of many-body physics on more general structures is unknown.
\par In this work we rectify this by approaching the many-body problem in an entirely new way. We take a generic spin $s$ Hamiltonian which encompasses a wide range of celebrated many-body models and treat the geometry as a parameter itself, encoding it in an underlying graph upon which the spins reside and interact via the edges. We then uncover the physics of the system when placing various levels of constraints on the graph. First, we place the minimal possible constraints and prove that in thermal equilibrium for a graph chosen uniformly at random from all possible simple graphs, almost surely, there is an absence of many-body physics and only collective, mean-field physics is possible. We achieve this by proving that, with increasing certainty and accuracy as the graph size $L$ increases, the free energy density on such a graph can be approximated with that for a single collective spin, as indicated in Fig. \ref{Fig:F1}.
\par Our result is based on the fact that, for the Erd\"{o}s-Renyi (ER) graph as $L \rightarrow \infty$, there exists no partition (cut) where the number of edges between the partitions differs significantly from its expected value. We then show that even for random graphs constrained to have a non-trivial cut the system can asymptotically almost surely (i.e. with probability tending to $1$ as $L \rightarrow \infty$) be reduced to a pair of interacting, large spins --- a result also depicted in Fig \ref{Fig:F1}. The emergence of complex, many-body physics is thus dependent on exceptional, highly structured graphs which are strongly distinct from almost all other graphs. We discuss how the sparse, regular lattices which commonly arise in nature are such an exception and go on to discover a hitherto unknown class of exceptional graphs which violate our proofs and where complex, many-body physics emerges: irregular dense graphs.

\par We illustrate these results. Taking the limit of our Hamiltonian which involves the competition between anti-ferromagnetic and XY couplings, we demonstrate how the nature of the underlying phase transition changes from non-existent, to first-order, to second-order, when considering random graphs chosen from the space of all graphs, all graphs with a non-trivial cut and dense inhomogeneous graphs, respectively. Using a well-established measure from image classification we demonstrate how, for the irregular dense graphs, the second-order phase transition coincides with significant complexity in images of the off-diagonal correlations in the system. Such a feature, which is not specific to the limit taken on our Hamiltonian, highlights the uniqueness of the ground-state on these structures and cannot occur in sparse, translationally-invariant structures. Our work here establishes the fate of many-body physics on a wide range of graphs and uncovers a new class of structures where novel, non-collective states of matter can emerge. 
\section*{Results}
\subsection*{Model and Hamiltonian}
We start by defining an arbitrary simple graph via $\mathcal{G} = \mathcal{G}(V,E)$ where $V$ are the $L = \vert V \vert$ vertices and $E$ are the $N_{E} = \vert E \vert$ edges. On each vertex of the graph we place a spin $s$ particle and have these spins interact with each other via the unweighted edges of the graph and be affected by a global field. The Hamiltonian for the total energy reads
\begin{equation}
    \hat{H}(\mathcal{G}) = \frac{L}{N_{E}}\left(\sum_{(v, v') \in E}\hat{h}_{v,v'}\right) + \sum_{v \in V}\hat{h}_{v},
    \label{Eq:GeneralHamiltonian}
\end{equation}
with $\hat{h}_{v,v'}$ and $\hat{h}_{v}$ being, respectively, one and two-body operators acting on the subscripted vertices. We take $\hat{h}_{v,v'} = J_{x}\hat{s}^{x}_{v}\hat{s}^{x}_{v'} + J_{y}\hat{s}^{y}_{v}\hat{s}^{y}_{v'} + J_{z}\hat{s}^{z}_{v}\hat{s}^{z}_{v'}$
where $J_{x}, J_{y}, J_{z} \in \mathbb{R}$. The canonical spin operator $\hat{s}^{\alpha}_{v}$ acts on spin $V$ in the $\alpha = x,y$ or $z$ direction. We also set $\hat{h}_{v} = \Vec{w} \cdot \Vec{s}^{ \alpha}_{v}$ where $\Vec{w} = (w_{x}, w_{y}, w_{z})$ and  $\Vec{s}^{ \alpha}_{v} = (s^{x}_{v}, s^{y}_{v}, s^{z}_{v})$. The scaling we have applied to the first term in $\hat{H}(\mathcal{G})$ means that it's largest eigenvalue (by absolute value) will scale as $\mathcal{O}(L)$ and thus the energy per spin is always finite, independent of the choice of graph.
\par The Hamiltonian $\hat{H}(\mathcal{G})$ encompasses a range of notable models of quantum magnetism and is a valid descriptor of non-magnetic systems such as fermions in the strongly interacting limit \cite{HeisenbergHubbard1} or bosons with a maximum on-site occupancy \cite{XXZHardcoreBosons1}. Throughout this work we will supplement our analytical results with numerical calculations for the limit of $\hat{H}(\mathcal{G})$ which describes the competition between spin-spin correlations along the $z$ and $x-y$ spin-axes respectively, i.e. where $\Vec{w} = 0$ and $J_{x} = J_{y}$. We should emphasize that the coupling strengths in our Hamiltonian are isotropic, meaning our results do not apply to setups such as the SYK
model, where the individual strengths are random and thus anisotropic.
\subsection*{The Average Graph}
\par To be as general as possible we will assume nothing about our simple graph $\mathcal{G}$ other than its size and draw it uniformly at random from the space of all simple graphs with $L$ vertices. Such a process is equivalent to drawing the graph from the Erd\"{o}s-Renyi (ER) ensemble \cite{Erdos1} with edges appearing independently with probability $p = 1/2$. We will reference an instance of a graph from the ER ensemble with edge probability $p$ as $\mathcal{G}_{\rm ER}(p)$ and reference our Hamiltonian on this graph via $\hat{H}(\mathcal{G}_{\rm ER}(p))$. In the case $p = 1$ then $\mathcal{G}_{\rm ER}(p)$ is equivalent to the complete graph $\mathcal{G}_{\rm Complete}$ -- the simple, unweighted graph on $L$ vertices where all edges are present. 
\par In order to determine the equilibrium physics of $\hat{H}(\mathcal{G}_{\rm ER}(p))$ we focus on the structure of the two-body operators, $\sum_{(v,v' \in E)}\hat{s}^{\alpha}_{v}\hat{s}^{\alpha}_{v'}$  in $\hat{H}(\mathcal{G}_{\rm ER}(p))$. The problem of finding the eigenspectrum of these operators is equivalent to finding the number of edges cut for all possible partitions of the underlying graph into $2s+1$ sets of vertices. On an ER graph with non-vanishing $p$ there exist strict bounds on this quantity and it cannot deviate significantly from its expected value \cite{ErdosRenyiCuts1, ErdosRenyiCuts2}. In the Supplementary Information (SI) we utilise such observations to derive a strict bound on the maximum eigenvalue (by magnitude) of $\hat{H}(\mathcal{G}_{\rm ER}(p)) - \hat{H}(\mathcal{G}_{\rm Complete})$ and subsequently prove the following theorem
\begin{theorem}
Let $\mathcal{G}_{\rm ER}(p)$ be an instance of the Erd\"{o}s-Renyi graph with finite edge probability $0 < p \leq 1$ and $L$ vertices. Let $\mathcal{G}_{\rm Complete}$ be the complete graph on $L$ vertices. Define the free-energy density of a $d^{L} \times d^{L}$ matrix as $f(\hat{A}) = -\frac{1}{L\beta}\ln \left({\rm Tr}(e^{-\beta \hat{A}})\right)$, where $\beta \in \mathbb{R}_{\geq 0}$ is the inverse temperature.
Then, for a given spin $s$ and an arbitrary set of values for the microscopic parameters $\{J_{x}, J_{y}, J_{z}, w_{x}, w_{y}, w_{z}\}$, 
\begin{equation}
    \lim_{L\rightarrow \infty}f(\hat{H}(\mathcal{G}_{\rm ER}(p))) \equiv  \lim_{L\rightarrow \infty}f(\hat{H}(\mathcal{G}_{\rm Complete})) \qquad  \forall \beta \in \mathbb{R}_{\geq 0},
    \label{Eq:HamiltonianEquivalence}
\end{equation}
Moreover, for finite large $L$, we have $\vert f(\hat{H}(\mathcal{G}_{\rm ER}(p))) -  f(\hat{H}(\mathcal{G}_{\rm Complete})) \vert = \mathcal{O}(L^{-1/2}) \ \ \forall \beta \in \mathbb{R}_{\geq 0}$.
\label{Theorem:TH1}
\end{theorem}
From Theorem \ref{Theorem:TH1} it follows that all thermodynamic observables (i.e. ones that can be written as a function of the free energy density) are equivalent for the equilibrium states $\rho(\mathcal{G}_{\rm ER}(p))$ and $\rho(\mathcal{G}_{\rm Complete})$, with $\rho(\mathcal{G}) \propto \exp(-\beta \hat{H}(\mathcal{G}))$.  The finite-size corrections, dictate that, for a single draw of the ER graph, the difference between between such observables scales as $\mathcal{O}(L^{-1/2})$. Due to their decreasing nature these statistical fluctuations about the average can be, with high probability, ignored for large $L$ --- with the limit $L \rightarrow \infty$ of $\mathcal{G}_{\rm ER}(p)$ being a single graph known as the Rado graph \cite{RadoGraph1}. Thus when drawing a single graph from the space of all graphs (i.e. setting $p = 1/2$) the equilibrium properties of the system will, with increasing certainty and accuracy as $L$ increases, be equivalent to those of $\rho(\mathcal{G}_{\rm Complete})$. In the SI we also provide numerical calculations supporting the bound we derive on the maximum eigenvalue of $\hat{H}(\mathcal{G}_{\rm ER}(p)) - \hat{H}(\mathcal{G}_{\rm Complete})$. 
\par To the best of our knowledge, this is the first proof of the equivalence between the ER and complete graph free energy densities for the general, quantum Hamiltonian in Eq. (\ref{Eq:GeneralHamiltonian}). Whilst such an equivalence has been proven for the classical Ising model \cite{ClassicalIsingProof1}, our theorem encompasses this classical case and applies to a much broader range of Hamiltonians -- both classical and quantum. 
\par Theorem \ref{Theorem:TH1} leaves us with a remarkable conclusion: the equilibrium physics described by $\hat{H}(\mathcal{G})$, where $\mathcal{G}$ is sampled from the space of all simple graphs, is not many-body. This is because the free energy density as $L \rightarrow \infty$ is equivalent to that for a Hamiltonian built solely from the collective spin operators $\hat{S}^{\alpha} = \sum_{v}\hat{s}^{\alpha}_{v}$ with $\alpha = x,y$ or $z$. The eigenstates of such a Hamiltonian are collective, mean-field states of matter such as condensates and uniform product states.

\begin{figure}[!t]
    \centering
    \includegraphics[width =0.8\textwidth]{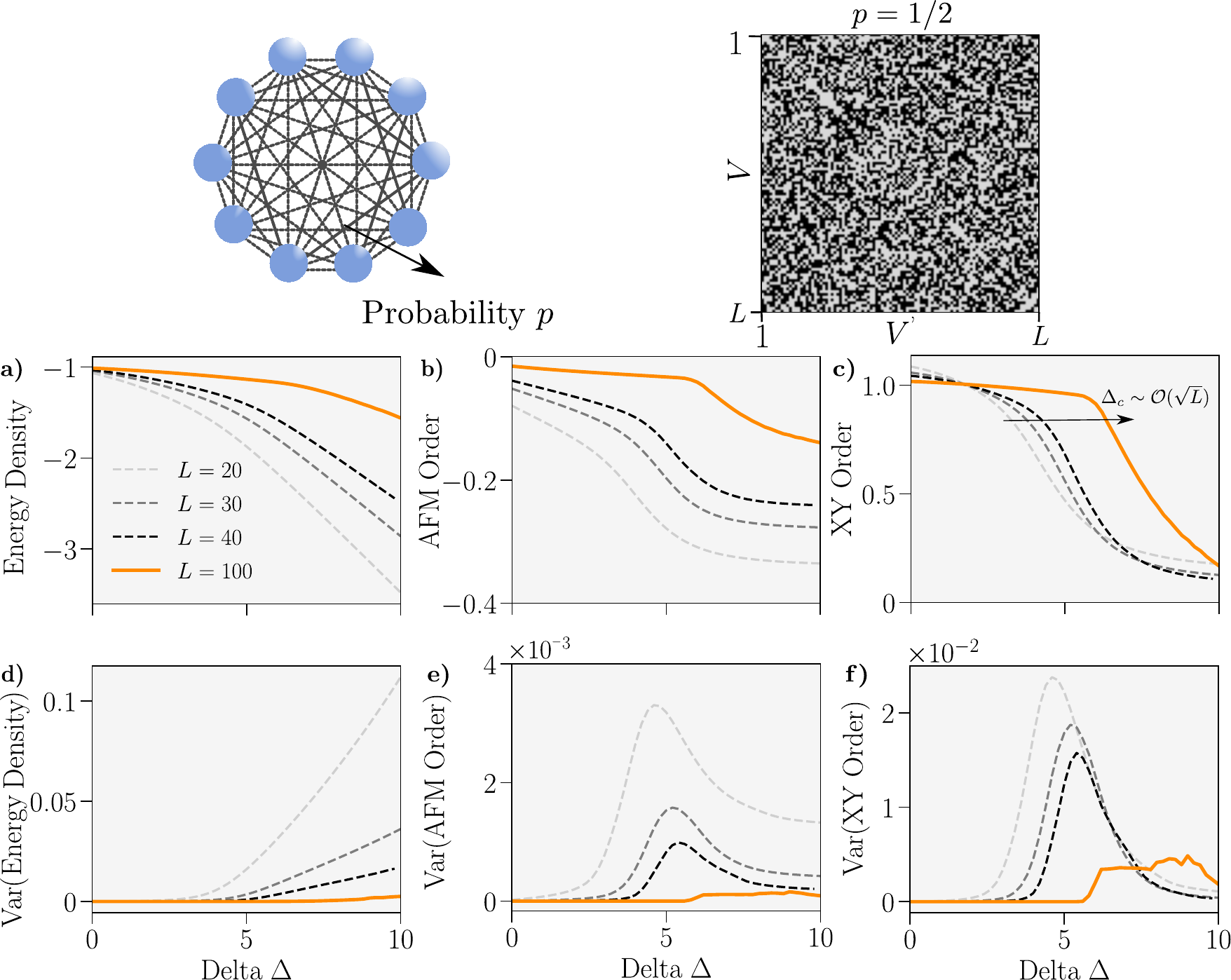}
    \caption{Properties of the ground-state of the spin $1/2$ XXZ Hamiltonian on the Erd\"{o}s-Renyi (ER) graph. A graph schematic is provided top left. We set $p=1/2$ here and an example adjacency matrix for $L= 100$ is shown for this parameter in the top right. Data is obtained, for a given $\Delta$, by calculating the ground state for $n$ of draws of the ER graph from its ensemble and then averaging (plots a-c) or taking the variance (plots d-f).  System sizes are coded by colour.
    \textbf{a-c)} Energy density $\frac{1}{L}\langle H \rangle$, anti-ferromagnetic order  $C_{\rm AFM}$ and XY order $C_{\rm XY}$ - see Eq. (\ref{Eq:OrderParameters}) -  versus $\Delta$. \textbf{d-f)} Variance of the corresponding upper observables. We used $n = 100, 100, 100$ and $10$ for $L = 20,30,40$ and $100$ respectively.}
    \label{Fig:F2}
\end{figure}

We reinforce these results in Fig. \ref{Fig:F2}, where we perform Matrix Product State (MPS) calculations of the ground-state of the spin $1/2$ XXZ Hamiltonian on the ER graph with $p = 1/2$. This Hamiltonian is explicitly defined in the Methods section, along with the order parameters $C_{\rm XY}$ and $C_{\rm AFM}$ for the XY and AFM (anti-ferromagnetic) phases repectively.
In the thermodynamic limit we can use Theorem \ref{Theorem:TH1} to identify the Dicke state (see SI) $\vert \psi \rangle_{\rm GS} = \vert L/2, M \rangle$ with $M$ finite as the ground state independent of the $z-z$ interaction strength $\Delta > 0$.
\par In Fig. \ref{Fig:F2} we observe, explicitly, the convergence, as $L$ increases, of the ground-state to this condensate for all values of $\Delta$ --- in agreement with Theorem \ref{Theorem:TH1}. Whilst on a finite-size system we observe a ground-state transition between the XY and AFM phases, we find that the critical point drifts and is disappearing as $L \rightarrow \infty$ due to an apparent scaling of $\Delta_{c} \sim \mathcal{O}(\sqrt{L})$ (see SI).  
We also calculate the variance of our observables (see Methods for definition) with respect to different draws of the graph from the ER ensemble. This variance decreases with $L$, indicating that the ground-state properties are converging to a single fixed limit for $L \rightarrow \infty$, where we can make statements about them with certainty. In the SI we use MPS methods to also demonstrate this convergence for the ground state of the Transverse Field Ising Model.
\par It is clear then that one must therefore pick exceptional graphs --- i.e. graphs from subspaces of the space of all graphs which are vanishingly small with respect to the full space --- in order to witness more complex, strongly-correlated states. These exceptions are graphs with some non-trivial cut where, even as $L \rightarrow \infty$, the number of edges between partitions deviates significantly from its expected value. We will now consider such graphs.
\begin{figure}[!t]
    \centering
    \includegraphics[width =0.8\textwidth]{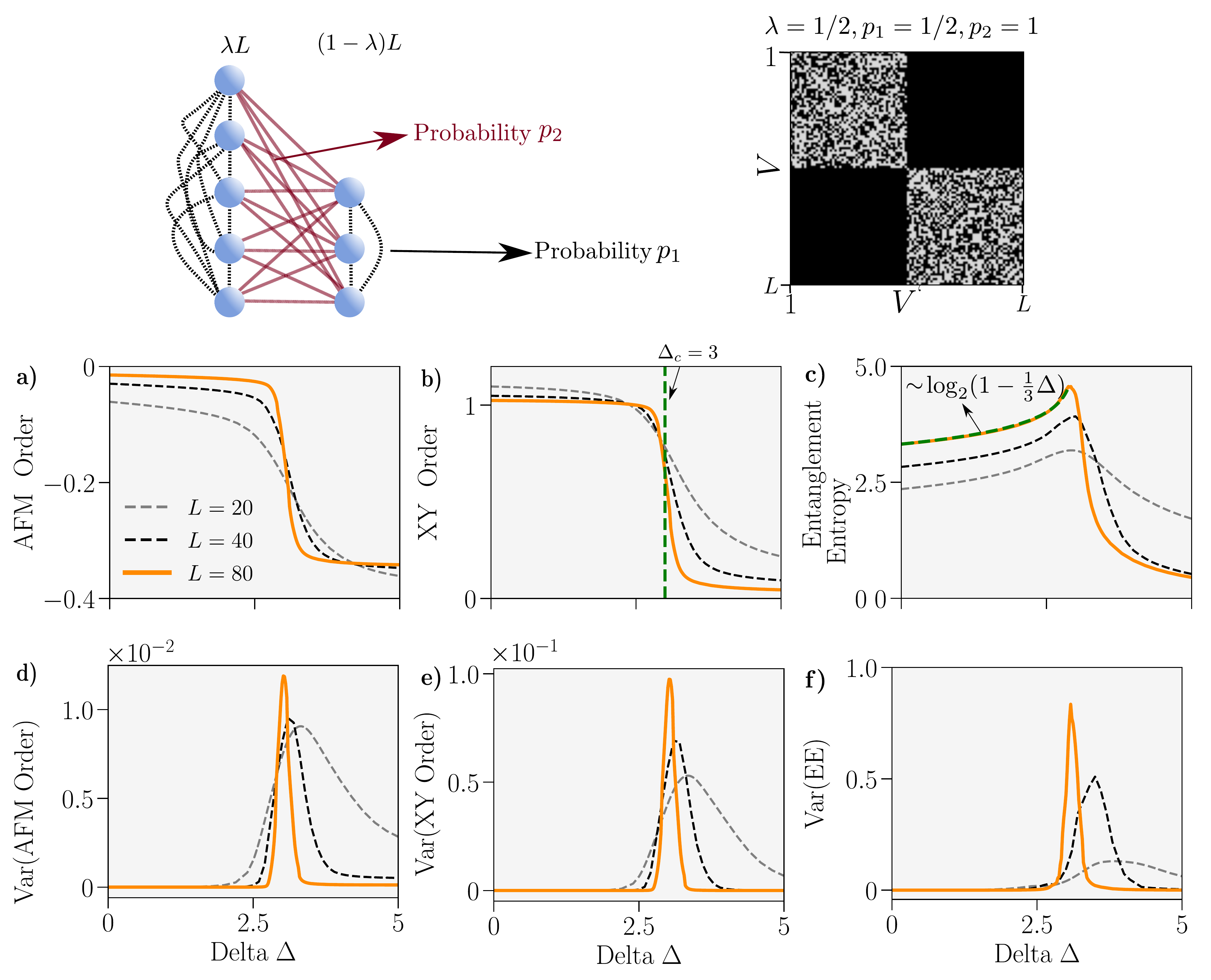}
    \caption{Properties of the ground-state of the spin $1/2$ XXZ Hamiltonian on a random graph with a non-trivial cut, $\mathcal{G}(\lambda, p_{1}, p_{2})$. A graph schematic is provided top left. We set $\lambda = 1/2$, $p_{2} = 1$ and $p_{2} = 1/2$ here and an example adjacency matrix for $L= 100$ and these parameters is shown top right.
    \textbf{a-c)} Average anti-ferromagnetic order  $C_{\rm AFM}$, XY order $C_{\rm XY}$ and Von-Neumann entanglement entropy (EE) between the two partitions versus $\Delta$ for graphs drawn $n$ times from $\mathcal{G}(\lambda, p_{1}, p_{2})$. A curve of the form $-0.265 \log_{2}(1-\frac{1}{3}\Delta) + {\rm EE}(\Delta = 0)$ is fitted to the entanglement entropy for $\Delta < 3$ and $L = 80$ and marked in green. The line $\Delta_{c} = 3$, which corresponds to the prediction of Eq. (\ref{Eq:CriticalPoint}), has been marked in green on the plot of the XY order. \textbf{d-f)} Variance of the corresponding upper observables. We used $n = 100$ for all values of $L$.}
    \label{Fig:F3}
\end{figure}
\subsection*{Graphs with a Non-trivial Cut}
\par We take a simple graph of $L$ sites along with a bi-partition of these sites into two sets $A$ and $B$ comprising $\lambda L$ and $(1-\lambda)L$ sites respectively ($0 < \lambda < 1/2$). We assume nothing other than that the ratio $\alpha$ of the number of edges between sites in different sets to the number of edges in the whole graph $N_{E}$  differs from its expected value of $2 \lambda (1-\lambda)$, even as $L \rightarrow \infty$. We reference a graph constructed uniformly at random from this ensemble as $\mathcal{G}(\lambda, p_{1}, p_{2})$, where $p_{1}$ and $p_{2}$ are finite numbers which can be directly related to $\alpha$ and $N_{E}(\mathcal{G}(\lambda, p_{1}, p_{2}))$ (see Methods).
\par In the SI we prove a theorem which dictates that the free energy density for $\hat{H}(\mathcal{G}(\lambda, p_{1}, p_{2}))$ is equivalent, in the thermodynamic limit, to that for an effective Hamiltonian which is built solely from the collective operators $\hat{S}^{\alpha}_{A} = \sum_{V \in A}\hat{s}^{\alpha}_{v}$ and $\hat{S}^{\alpha}_{B} = \sum_{V \in B}\hat{s}^{\alpha}_{v}$. This leads to another striking conclusion: if one were to randomly select a graph from those with a non-trivial cut they would find, as $L$ increases, that the system realised by $\hat{H}(\mathcal{G})$ is increasingly similar to that of a pair of interacting collective spins. Whilst the physics of such a system is richer than that for the ER ensemble, it is still collective and not many-body.
\par In Fig. \ref{Fig:F3} we apply our MPS calculations to the $s=1/2$ XXZ limit of $\hat{H}(\mathcal{G}(\lambda, p_{1}, p_{2}))$ in order to reinforce this result. We directly observe an emergent first-order transition between an XY phase and an AFM phase. The properties of the ground state in the XY regime demonstrate convergence towards the condensate $\ket{L/2, 0}$ whilst, in the AFM regime, the properties are that of the product state where the spins in the sets $A$ and $B$ polarise in opposite directions along the spin-$z$ axis. These results are in agreement with our reduction of the system to a pair of collective spins, which predicts (see SI) a critical point of 
\begin{equation}
    \Delta_{c} = \frac{p_{1}(\lambda^{2} + (1-\lambda)^{2}) + 2p_{2}\lambda(1-\lambda)}{2p_{2}\lambda(1-\lambda) - p_{1}(\lambda^{2} + (1-\lambda)^{2})}.
    \label{Eq:CriticalPoint}
\end{equation}
\par Again, we are averaging the ground-state properties over draws of the graph from its corresponding ensemble. With increasing $L$ the peak in the variance appears to be tending towards a delta function centred on $\Delta_{c}$, where we don't expect the variance to vanish due to the first-order discontinuity.

\begin{figure}[!t]
    \centering
    \includegraphics[width =0.8\textwidth]{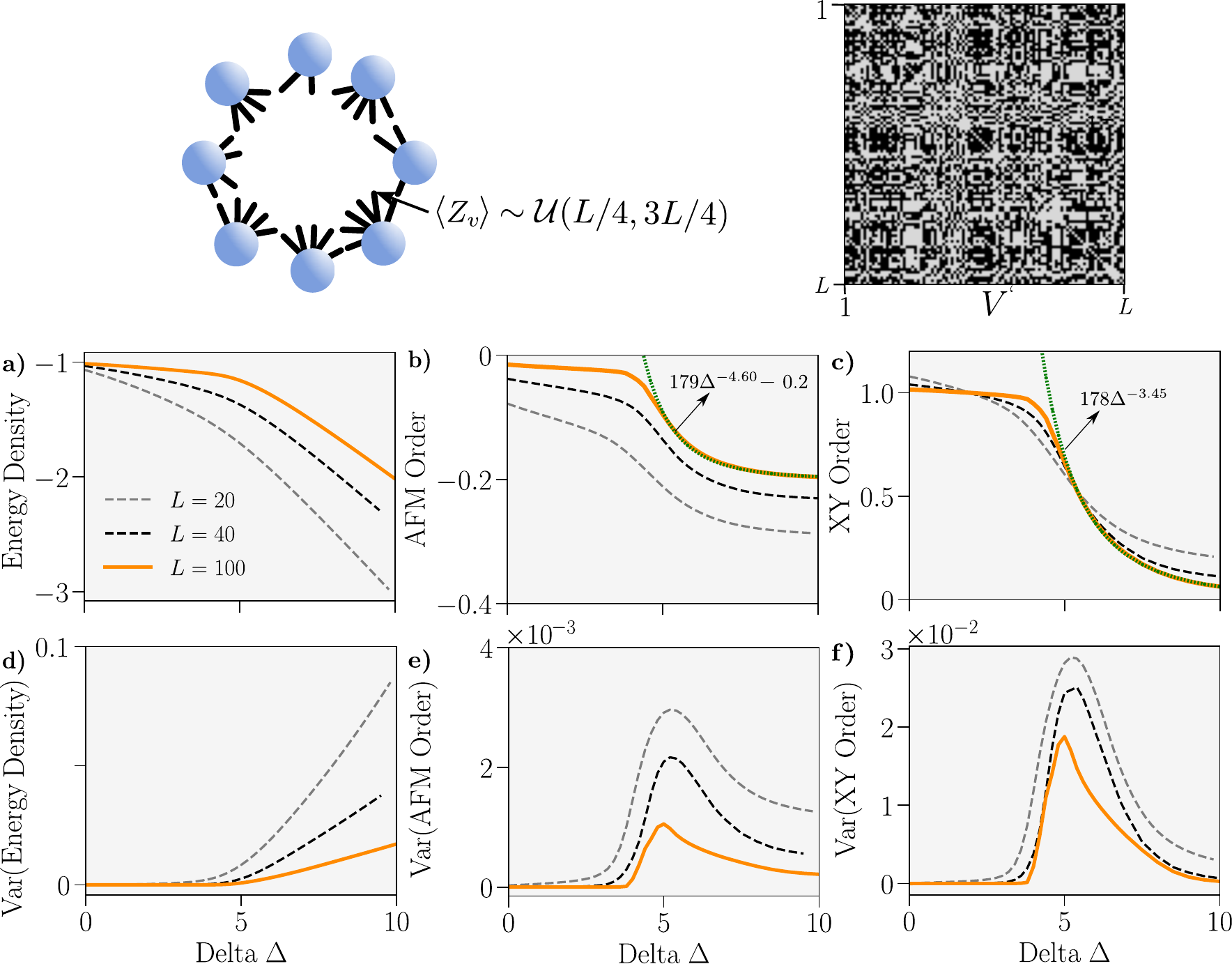}
    \caption{Properties of the ground-state of the spin $1/2$ XXZ Hamiltonian on the graph where each site has a degree $Z_{V}$ drawn from $\mathcal{U}(L/4, 3L/4)$, the uniform discrete distribution over the interval $[L/4, 3L/4]$. Graph schematic alongside adjacency matrix for $L= 100$ is provided. 
    \textbf{a-c)} Average Energy Density, anti-ferromagnetic order  $C_{\rm AFM}$ and XY order $C_{\rm XY}$ versus $\Delta$ for graphs drawn $n$ times from the ensemble. Power-law curves have been fitted to the XY and AFM orders for $\Delta > 2$ and marked in green. \textbf{d-f)} Variance of the corresponding upper observables. We used $n = 100, 100$ and $10$ for $L = 20, 40$ and $100$ respectively.}
    \label{Fig:F4}
\end{figure}

\par What is the key property of the graphs $\mathcal{G}_{\rm ER}(p)$ and $\mathcal{G}(\lambda, p_{1}, p_{2})$ which allow the equilibrium physics of $\hat{H}(\mathcal{G})$ to be reducible to a collective spin model? Despite the fact they are not invariant under a given permutation of two spins, our proof tells us that the statistical fluctuations which break these symmetry fluctuations cannot affect the thermodynamic properties of the system. As a result we can ignore them as $L \rightarrow \infty$ and worth with a simpler Hamiltonian built solely from collective spin operators. The equilibrium properties of the system can then be calculated with a semi-classical solution. It is worth emphasizing that such a solution does not capture the quantum fluctuations
which are included in our numerics and present in the true ground state in the form of non-zero entanglement entropy and simultaneous long-range order along the $x$ and $y$ spin axes. 
\par These results necessitate searching for exceptional structures where complex, truly many-body physics is manifest. Sparse, regular structures like the hypercubic lattice are the well-known exception as our proofs are reliant on the graph being dense (i.e. $N_{E}/L^{2}$ cannot vanish as $L \rightarrow \infty$). Theoretical results for various limits of $\hat{H}(\mathcal{G})$ on low-dimensional instances of such structures are numerous and their capacity to host complex, many-body states of matter is well established \cite{EstablishedPhaseDiagram1, EstablishedPhaseDiagram2, EstablishedPhaseDiagram3, EstablishedPhaseDiagram4, EstablishedPhaseDiagram5}. Despite being so strongly distinct from the average case, such structures are ubiquitous in nature. Our results suggest that if these exceptional structures were not commonplace, the world around us would not be able to exhibit such complex, rich behaviour. 
\subsection*{Irregular Dense Graphs}
\par Importantly, even within the space of dense graphs we are able to discover hitherto unknown exceptional structures which can host complex many-body phases of matter. As far as we are aware, such structures have never before been treated in the realm of many-body physics. Specifically, in the following, we consider `irregular dense graphs' where, even as $L \rightarrow \infty$, we cannot split the sites into a finite number of sets where sites in the same set have identical values of $Z_{V}/L$, with $Z_{V}$ the co-ordination number of a given site $V$. Such a statement is not true of $\mathcal{G}_{\rm ER}(p)$ and $\mathcal{G}(\lambda, p_{1}, p_{2})$.

\begin{figure}[!t]
    \centering
    \includegraphics[width =0.8\textwidth]{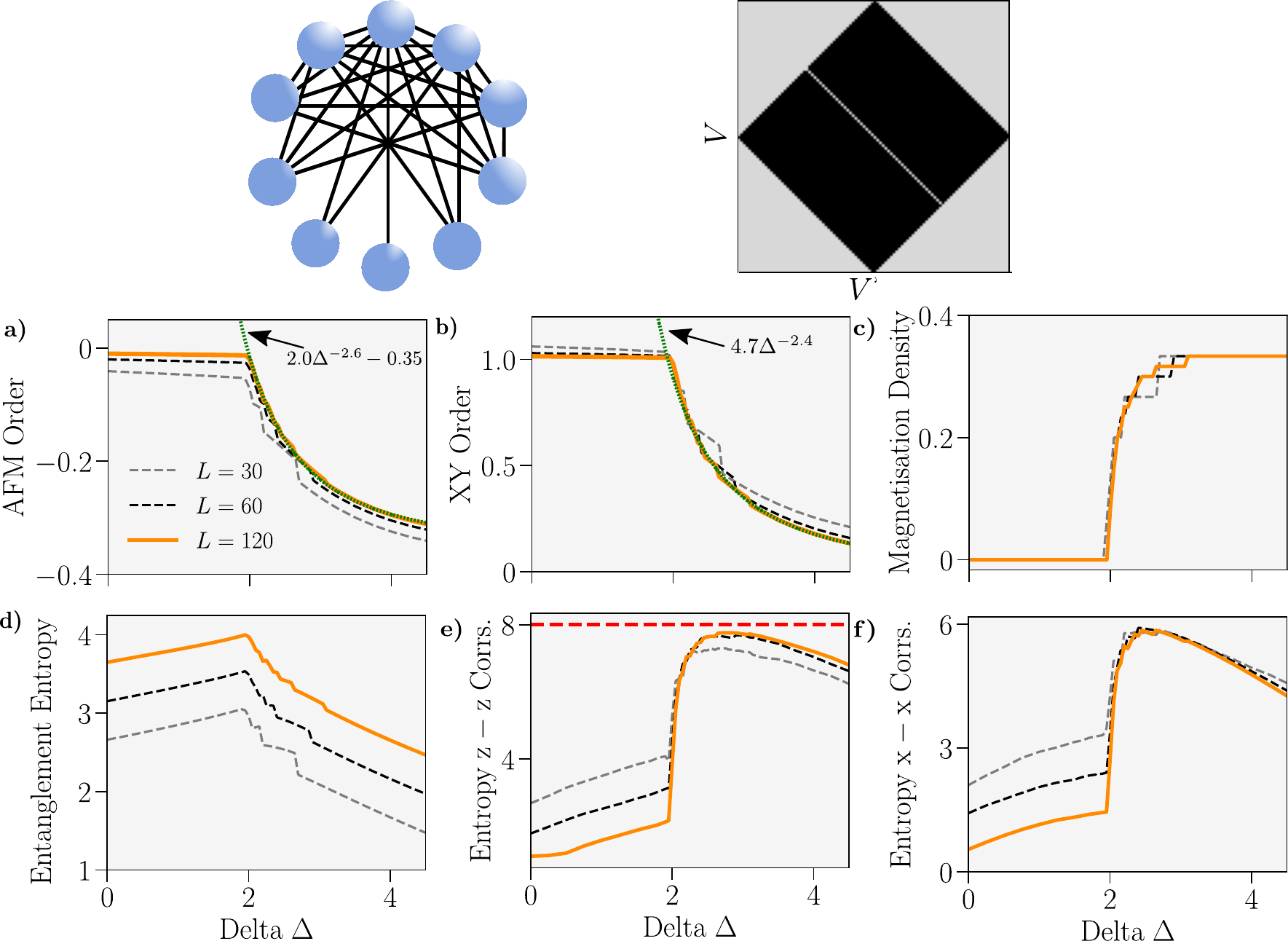}
    \caption{Properties of the ground-state of the spin $1/2$ XXZ Hamiltonian on the `Maximally Irregular Graph' where each pair of vertices, other than two, have a different degree. Graph schematic alongside adjacency matrix for $L= 100$ is provided. System sizes are coded by colour.
    \textbf{a-f)} Anti-ferromagnetic order $C_{\rm AFM}$, XY order $C_{\rm XY}$, magnetisation density $\langle \hat{S}^{z} \rangle / L$, von-Neumann entanglement entropy between sites $V = 1 ... L/2$ and $V = L/2 + 1 ... L$ and Shannon-Entropy of the $z-z$ and $x-x$ correlations (see Methods) versus $\Delta$. We use $n = 256$ bins to calculate the Shannon Entropy. Power-law curves have been fitted to $C_{\rm AFM}$ and $C_{\rm XY}$ for $\Delta > 2$ and $L = 120$ and marked in green. A dotted red line has been added to plot e) to indicate the maximum possible value for the Shannon Entropy with the numbers of bins used.}
    \label{Fig:F5}
\end{figure}

\par As a prototypical example of an irregular dense graph we consider one drawn from the ensemble of graphs where the degree of each site is assigned a value uniformly in the range $[L/4, 3L/4]$. Details on how we construct an instance of this graph are given in the Methods. We perform MPS calculations to find the ground-state of the XXZ model on this inhomogeneous structure, with the results pictured in Fig. \ref{Fig:F4}. Again, the general reduction in the variance of observables as $L$ increases suggests that there is a well-defined thermodynamic limit for this ensemble, with an emergent continuous phase transition between the XY and AFM phases. We find, by numerical fitting, the associated order parameters follow a power law for $\Delta > \Delta_{c} \approx 2$. 
\par We expand on these results further by considering, in Fig. \ref{Fig:F5}, the `maximally irregular' graph where every site, other than one pair of sites by necessity, has a different degree. We observe an emergent second-order phase transition on this graph with the same functional scaling of the order parameters as in Fig. \ref{Fig:F4}. Unlike the other graphs considered in this work, the ground-state of the system does not always possess $\langle \hat{S}^{z} \rangle = 0$ and its variation with $\Delta$ is responsible for the discrete changes in the observables in Fig. \ref{Fig:F5}.
\par As a measure of the complex, inhomogeneous nature of the ground-state on the maximally irregular graph we calculate the Shannon Entropy (SE) of the matrices of two-point $z-z$ and $x-x$ correlations (see Methods for definition) as a function of $\Delta$. These matrices can be viewed as images of the correlations in the system with each two-point correlation constituting a `pixel' whose value is bounded between $1$ and $-1$. The SE of this image then specifies the amount of information contained within it, and is a widely used measure in image reconstruction and classification algorithms \cite{ImageEntropy2, ImageEntropy3}. Here we apply this measure to the image of correlations in a quantum system, noting that it vanishes if all two-point correlations are the same (such as in the Dicke-state $\ket{L/2, M}$) and is maximised when the correlations are distributed uniformly across $[-1,1]$.
\par In Fig. \ref{Fig:F5} we observe a sharp increase in the SE for the $x$ and $z$ correlations at the critical point, with the $z-z$ correlations close to maximising this entropy and meaning that the corresponding correlation matrix cannot be compressed without a loss of information \cite{LosslessTheorem1}. The phase transition in the maximally irregular graph is thus heralded by a vast amount of complexity in the system's correlations and along all spin-axes. In the SI we compare this entropy for the graphs considered in this work as well as the 1D chain. The entropy is largest for the irregular dense graphs and these are the only ones where it does not diminish with system size and is non-zero simultaneously in the $x$, $y$ and $z$ degrees of freedom as $L \rightarrow \infty$. We also perform calculations for the Transverse Field Ising limit of $\hat{H}(\mathcal{G})$ on the maximally irregular graph and observe similar behaviour in this entropy as $L \rightarrow \infty$ in the vicinity of the critical point. 
\par Unlike other existing measures of complexity in quantum states such as the average disparity \cite{Complexity2}, features like translational invariance are accounted for by our measure and reduce the informational complexity associated with a quantum state. Given that there is no disorder or anisotropy in the microscopic parameters of our Hamiltonian, such results warrant our interpretation of irregular dense graphs as a new class of geometries on which many-body systems can exhibit novel, non-trivial phases of matter.
\section*{Discussion}
\par Our results open up a number of promising avenues for future research. Numerical methods in Condensed Matter physics have been optimised over the last few decades to treat quantum systems hosted on regular, sparse, translationally-invariant structures. Our work instigates the need to alter existing methods, in order to  account for more complex, inhomogeneous geometries and allow us to probe the physics, both in and out-of-equilibrium, of such systems.
\par In an equilibrium context, it would be important to understand how geometric irregularity affects the properties of various Hamiltonians -- including those with anisotropic, random couplings \cite{SYKSpinGlass1}. In an out-of-equilibrium setting, a number of questions arise about how exotic phenomena such as time-crystalline order \cite{TimeCrystals1, TimeCrystals2}, quantum synchronisation \cite{QuantSynchronisation1, QuantSynchronisation2}, heating-induced order \cite{HeatingInducedOrder2} are affected in the presence of such geometric irregularity. These results could enable the engineering of quantum states of matter in many-body simulators with new, geometrically enabled functionalities.

\section*{Methods}\label{sec11}
For our numerical results we used state-of-the-art Matrix Product State calculations for the XXZ $s = 1/2$ limit of Eq. (\ref{Eq:GeneralHamiltonian}). The Hamiltonian in this limit reads
\begin{equation}
    H_{\rm XXZ}(\mathcal{G}) = \frac{L}{N_{E}}\sum_{v,v' \in E}-J(\hat{\sigma}_{v}^{x}\hat{\sigma}^{x}_{v'} + \hat{\sigma}_{v}^{y}\hat{\sigma}^{y}_{v'}) + \Delta \hat{\sigma}^{z}_{v}\hat{\sigma}^{z}_{v'},
    \label{Eq:XXZHamiltonian}
\end{equation}
which corresponds to setting, in Eq. (\ref{Eq:GeneralHamiltonian}), $J_{x} = J_{y} = -J$, $J_{z} = \Delta$, $\Vec{w} = 0$ and $\hat{s}^{\alpha}_{v} = \hat{\sigma}^{\alpha}_{v}$ --- dropping the factor of $1/2$ in front of the Pauli matrix $\hat{\sigma}^{\alpha}_{v}$ for simplicity. We restrict ourselves to $\Delta \geq 0$, set $J = 1$ and define order parameters for the XY and anti-ferromagnetic (AFM) phases respectively via
\begin{align}
    &C_{\rm AFM} = \frac{1}{N_{E}}\sum_{(v,v') \in E } \langle \hat{\sigma}^{z}_{v}\hat{\sigma}^{z}_{v'} \rangle , \notag \\
    &C_{\rm XY} = \frac{1}{N_{E}}\sum_{\substack{v,v'=1 \\ v > v'}}^{L} \langle \hat{\sigma}^{x}_{v}\hat{\sigma}^{x}_{v'}+\hat{\sigma}^{y}_{v}\hat{\sigma}^{y}_{v'} \rangle.
    \label{Eq:OrderParameters}
\end{align}
These take a non-zero value in their respective phases, vanish in the opposing phase and, importantly, can be calculated for the ground state on any graph. The XXZ Hamiltonian has rotational symmetry around the spin-$z$ axis meaning $\langle \hat{\sigma}^{x}_{v}\hat{\sigma}^{x}_{v} \rangle = \langle \hat{\sigma}^{y}_{v}\hat{\sigma}^{y}_{v} \rangle$ for the ground state. The parameter $C_{\rm XY}$ thus quantifies the simultaneous off-diagonal order in the $x$ and $y$ degrees of freedom.  
\par In this work we also introduce the Shannon Entropy of the spin-correlations in a given state as
\begin{equation}
    H\left(\langle \hat{\sigma}^{\alpha}_{v}\hat{\sigma}^{\alpha}_{v'}\rangle \right) = \sum_{i = 0}^{n-1}p_{i}\log_{2}(p_{i}),
\end{equation}
where $p_{i}$ is the fraction of elements of the $L \times L$ matrix $\langle \hat{\sigma}^{\alpha}_{v} \hat{\sigma}^{\alpha}_{v'}\rangle$ which are between $-1 + 2i/n$ and $-1 + 2(i+1)/n$. The integer $n$ is the number of bins used to `bin up' the matrix elements. We use $n = 256$ throughout in order to make the connection between $\langle \hat{\sigma}^{\alpha}_{v} \hat{\sigma}^{\alpha}_{v'}\rangle$ and a grayscale image of the correlations in the system. This entropy measure is bounded as $0 \leq H(\langle \hat{\sigma}^{\alpha}_{v}\hat{\sigma}^{\alpha}_{v'}\rangle) \leq \log_{2}(n)$ and can be interpreted as the amount of information required to encode the distribution or image of off-diagonal correlations in a state.
\par Our DMRG calculations often involve ensemble-averaging the ground state properties over a series of random graphs drawn from a corresponding ensemble for a given $\Delta$ and $L$. In order to assess the convergence of the order parameters $C_{\rm XY}$ and $C_{\rm AFM}$ to a well defined thermodynamic limit we calculate their variances ${\rm Var}(C_{\rm XY})$ and ${\rm Var}(C_{\rm AFM})$. These are defined as
\begin{equation}
    {\rm Var}(C) = \frac{1}{n-1}\sum_{i=1}^{n}(C^{i} - \bar{C})^{2},
\end{equation}
for $n$ draws of the graph from its ensemble and where $\bar{C}$ is the average of the ground state order parameter for these different draws. For a given $L$ and $\Delta$, the variance then tells us the fluctuation in the ground state properties when averaging over random instances of the given graph ensemble. As the order parameters have been appropriately normalised by the size of the graph the variance will allow us to infer the convergence of the ground state properties in the thermodynamic limit. We discuss explicit details of our Matrix Product State calculations and implementation in the Supplementary Information. We also provide analysis of the truncation errors and energy convergence involved in our DMRG calculations. The ground-state in our calculations typically has a bi-partite entanglement entropy that scales logarithmically with the partition size and makes it tractable using a bond-dimension proportional to the system size. Such a scaling has previously been observed in the Lipkin-Meshov-Glick model \cite{LMGModel1} (an all-to-all XY model with a transverse field) and here we observe it for the XXZ model on a range of dense graphs.
\par In the main text we introduced the non-trivial cut graph where we assumed nothing but that there exists some bi-partition of the $L$ vertices into two sets, size $\lambda L$ and $(1-\lambda)L$, and where the cut-size $\alpha$ (i.e. the ratio of the number of edges between the two sets to the total number of edges in the graph $N_{E}$) is not equal to its expected value $2 \lambda (1-\lambda)$. We can construct such a graph by taking $L$ sites, partitioning them into the two corresponding sets and randomly assigning edges between sites in the same set with probability $p_{1}$ and between sites in different sets with probability $p_{2}$. The values of $\alpha$ and $N_{E}$ are then approximately
\begin{align}
    &N_{E} \approx p_{1}(\lambda^{2} -\lambda + 1/2) + p_{2}\lambda(1-\lambda)L^{2}, \notag \\
    &\alpha \approx \frac{2p_{2}\lambda(1-\lambda)}{p_{1}\lambda^{2} + p_{1}(1-\lambda)^{2} + 2p_{2}\lambda(1-\lambda)},
\end{align}
which becomes exact as $L \rightarrow \infty$, with each unique pair of values of $p_{1}$ and $p_{2}$ uniquely specifying $N_{E}$ and $\alpha$. As a result, the parameters $p_{1}$ and $p_{2}$ can be interchanged freely with $\alpha$ and $N_{E}$ in this limit and lead us to define an instance of this graph as $\mathcal{G}(\lambda, p_{1}, p_{2})$. Provided $p_{1} \neq p_{2}$ and $0 < \lambda < 1$ then we find $\alpha \neq 2 \lambda (1-\lambda)$, meaning $\mathcal{G}(\lambda, p_{1}, p_{2})$ has a non-trivial cut. Our construction routine --- given its independent treatment of edges --- will uniformly sample over all graphs with $L$ vertices, a number of edges $N_{E}$ and non-trivial cut-size $\alpha$.
\par We also introduced the uniform variate graph where the degree of each site follows the discrete uniform distribution $\mathcal{U}(L/4, 3L/4)$. We have used lower and upper bounds sufficiently separated from $0$ and $L - 1$ to avoid creating non-graphical degree
distributions. To generate a given graph from this ensemble, we draw the degree distribution by randomly generating the degree of each site --- repeating until sum of the degrees is even --- and then using the Havel Hakimi (HH) \cite{HavelHakimi} algorithm to generate a graph with the given degree distribution. Such a routine does not sample uniformly from the space of all graphs with degrees drawn from the distribution $\mathcal{U}(L/4, 3L/4)$ due to the high assortativity \cite{Assortativity1} bias in the HH algoritithm. We are, however, unaware of an algorithm which can generate unbiased samples of dense, inhomogeneous graphs with a given degree distribution. Hence, our results are for the ensemble of graphs with degrees drawn from the distribution $\mathcal{U}(L/4, 3L/4)$ and generated by the HH algorithm.

\section*{Data Availablity}
The data that was used to create the plots within this paper is provided as source data.

\section*{Code Availability}
The Network Python (TeNPy) library \cite{TENPY1}, which can be used to perform the simulations in the article, is available at https://tenpy.readthedocs.io/en/latest/. The programming scripts used to obtain the source data in this manuscript are available from the corresponding author upon reasonable request.

\section*{Acknowledgements}
The simulations used to produce the computational results of our work were run
on the University of Oxford Advanced Research Computing (ARC) facility and involved over over $100,000$ hours of exclusive use of nodes each equipped a with 48 core 2.9Ghz Intel Cascade Lake Processor. JT is grateful for ongoing support through the Flatiron Institute, a division of the Simons Foundation. DJ and JT acknowledge funding from EPSRC grant EP/P009565/1. DJ also acknowledges funding from the Cluster of Excellence ‘Advanced Imaging of Matter’ of the Deutsche Forschungsgemeinschaft (DFG) - EXC 2056 - project ID 390715994. AS acknowledges funding from the UK Engineering and Physical Sciences Research Council as well as from the Smith-Westlike scholarship.

\section*{Author Contributions}
JT provided the idea for the project, formulated the proofs and wrote the manuscript. JT, AS and AA ran the numerical simulations and analysed the data. JT and DJ edited the manuscript with help from AS and AA. All authors contributed substantially to discussions on the results and ideas contained therein. DJ oversaw and managed the overall project.

\section*{Competing Interests}
There are no competing interests.

\newpage
\section*{\fontsize{24}{15}\selectfont Supplementary Information for Quantum physics in Connected Worlds}
\renewcommand{\theequation}{S\arabic{equation}}
\renewcommand{\figurename}{Supplementary Figure}
\setcounter{figure}{0}   
\setcounter{equation}{0}   

\renewcommand{\thetheorem}{S\arabic{theorem}}
\renewcommand{\thelemma}{S\arabic{lemma}}
\renewcommand{\thesubsection}{Section}

\section*{Proof of the Equivalence of the XYZ Free Energy on the Complete and Erd\H{o}s-R\'enyi (ER) Graphs}
We redefine, for clarity, the Hamiltonian from the main text
\begin{equation}
    \hat{H}(\mathcal{G}) = \frac{L}{N_{E}}\left(\sum_{(v, v') \in E}\hat{h}_{v,v'}\right) + \sum_{v \in V}\hat{h}_{v},
    \label{Eq:SMGeneralHamiltonian}
\end{equation}
with
\begin{align}
    &\hat{h}_{v,v'} = J_{x}\hat{s}^{x}_{v}\hat{s}^{x}_{v'} + J_{y}\hat{s}^{y}_{v}\hat{s}^{y}_{v'} + J_{z}\hat{s}^{z}_{v}\hat{s}^{z}_{v'}, \qquad J_{x},J_{y}, J_{z} \in \mathbb{R}, \notag \\
    &\hat{h}_{v} = w_{x}\hat{s}_{v}^{x} + w_{y}\hat{s}_{v}^{y} + w_{z}\hat{s}^{z}_{v}, \qquad w_{x},w_{y}, w_{z} \in \mathbb{R}.
    \label{Eq:SMGeneralHamiltonianParameters}
\end{align}
All quantities retain their meaning from the main text. We also reintroduce the free energy density of a matrix of size $d^{L} \times d^{L}$ as
\begin{equation}
    f(\hat{A}) = -\frac{1}{L\beta}\ln \left({\rm Tr}(e^{-\beta \hat{A}})\right), \qquad \beta \in \mathbb{R}_{\geq 0}.
    \label{Eq:SMFreeEnergy}
\end{equation}
In this section we prove the following theorem from the main text:
\begin{theorem}
Let $\mathcal{G}_{\rm ER}(p)$ be an instance of the Erd\H{o}s-R\'enyi graph with finite edge probability $0 < p \leq 1$ and $L$ vertices. Let $\mathcal{G}_{\rm Complete}$ be the complete graph over $L$ vertices. Then, given an arbitrary set of values for the microscopic parameters $\{J_{x}, J_{y}, J_{z}, w_{x}, w_{y}, w_{z}\}$, 
\begin{equation}
    \lim_{L\rightarrow \infty}f(\hat{H}(\mathcal{G}_{\rm ER}(p))) = \lim_{L\rightarrow \infty}f(\hat{H}(\mathcal{G}_{\rm Complete})) \qquad \forall \beta \in \mathbb{R}_{\geq 0}.
    \label{Eq:SMFreeEnergyEquivalence}
\end{equation}
Moreover, for finite $L$, we have $\vert f(\hat{H}(\mathcal{G}_{\rm ER}(p))) -  f(\hat{H}(\mathcal{G}_{\rm Complete})) \vert = \mathcal{O}(L^{-1/2})$.
\label{Theorem:SMTH1}
\end{theorem}

\par Theorem \ref{Theorem:SMTH1} implies the equivalence, at any temperature, between the equilibrium states (with trace unity) $\rho(\mathcal{G}_{\rm ER}(p)) \propto \exp(-\beta \hat{H}(\mathcal{G}_{\rm ER}(p)))$ and $\rho(\mathcal{G}_{\rm Complete}) \propto \exp(-\beta \hat{H}(\mathcal{G}_{\rm Complete}))$ in terms of any thermodynamic quantity which can be written as a function of the free energy density. Such quantities dictate the macroscopic phase of the system and are thus of great import. We do not, however, make statements about out-of-equilibrium scenarios or purely local observables which cannot be derived from the free energy density --- these are beyond the scope of this paper and an interesting area of future study. 

As a high level sketch of the proof: We will prove a Lemma on a pair of Hermitian matrices $\hat{A}$ and $\hat{B}$ which dictates that the difference in their free energy densities can be bounded in terms of the largest (by magnitude) eigenvalue of $\hat{A}-\hat{B}$.  We will then, first for spin $s=1/2$ and then generally for finite $s$, explicitly identify the scaling on the largest eigenvalue (by magnitude) of $ \hat{H}(\mathcal{G}_{\rm ER}(p)) - \hat{H}(\mathcal{G}_{\rm Complete}) $ and invoke the aforementioned Lemma.
\par The Lemma that we will use to help prove Theorem \ref{Theorem:SMTH1} is as follows
\begin{lemma}
    Let $\hat{A}_{1}, \hat{A}_{2}, ... $ and $\hat{B}_{1}, \hat{B}_{2}, ... $ be two sequences of Hermitian matrices. The matrices $A_{L}$ and $B_{L}$ in the sequence have size $d^{L} \times d^{L}$ -- with $d$ fixed. Let $\hat{D}_{L} = \hat{A}_{L} - \hat{B}_{L}$ and $\lambda^{D}_{\rm Max}$ be the largest (in terms of absolute value) eigenvalue of $\hat{D}_{L}$. If $\vert \lambda^{D}_{\rm Max} \vert = \mathcal{O}(L^{\gamma})$ then $\vert f(\hat{A}_{L}) - f(\hat{B}_{L}) \vert = \mathcal{O}(L^{\gamma - 1})$, which vanishes for $\gamma < 1$ as $L \rightarrow \infty$.
    \label{Lemma:SMLemma1}
\end{lemma}
We proceed to prove this Lemma, dropping the subscript on $\hat{A}_{L}$ and $\hat{B}_{L}$ -- from here on it is always implied that they are of size $d^{L} \times d^{L}$. First, order the eigenvalues of $\hat{A}$, $\hat{B}$ and $\hat{D}$ as $\lambda^{\eta}_{1}, \lambda^{\eta}_{2}, ..., \lambda^{\eta}_{d^{L}}$, with $\lambda^{\eta}_{i} \geq \lambda^{\eta}_{i-1}$ and $\eta = \hat{A},\hat{B}$ or $\hat{D}$ . By Weyl's inequality we have $\lambda^{B}_{i} + \lambda^{D}_{1} \leq \lambda^{A}_{i} \leq \lambda^{B}_{i} + \lambda^{D}_{d^{L}}$. Now, it follows from this inequality and $\vert \lambda^{D}_{\rm Max} \vert = \mathcal{O}(L^{\gamma})$ that $\vert \lambda^{A}_{i} - \lambda^{B}_{i} \vert = \mathcal{O}(L^{\gamma}) \ \forall i$. We can therefore write $\lambda^{A}_{i} = \lambda^{B}_{i} + c_{i}$ where $c_{i}$ is a real number such that $\vert c_{i} \vert  = \mathcal{O}(L^{\gamma})$.
\par Now define the absolute free energy density difference of $\hat{A}$ and $\hat{B}$, substituting in $\lambda^{A}_{i} = \lambda^{B}_{i} + c_{i}$:
\begin{equation}
    \Delta f = \vert f(\hat{A}) - f(\hat{B}) \vert = \Big\vert \frac{1}{L}\ln \left( \frac{{\rm Tr}(\exp(-\beta \hat{A}))}{{\rm Tr}(\exp(-\beta \hat{B}))} \right) \Big\vert = \Big\vert \frac{1}{L}\ln \left( \frac{\sum_{i=1}^{d^{L}}\exp(-\beta(c_{i}))\exp(-\beta(\lambda^{A}_{i}))}{\sum_{i=1}^{d^{L}}\exp(-\beta(\lambda^{A}_{i}))} \right) \Big\vert.
\end{equation}
Define the following vectors of dimension $d^{L}$:
\begin{align}
    &\mathbf{v} = \left(\exp(-\beta\lambda^{A}_{1}/2), \exp(-\beta\lambda^{A}_{2}/2), ..., \exp(-\beta\lambda^{A}_{d^{L}}/2)\right)^{T}, \notag \\ &\mathbf{c} = \left(\exp(-\beta c_{1}), \exp(-\beta c_{2}), ..., \exp(-\beta c_{d^{L}})\right)^{T},
\end{align}
and the matrix $\hat{C} = {\rm Diag}(\mathbf{c})$. Observe that
\begin{equation}
    \Delta f = \Big\vert \frac{1}{L}\ln \left( \frac{\langle \mathbf{v} \vert \hat{C} \vert \mathbf{v} \rangle}{\langle \mathbf{v} \vert \mathbf{v} \rangle} \right) \Big\vert = \Big\vert \frac{1}{L}\ln \left( \langle \tilde{\mathbf{v}} \vert \hat{C} \vert \tilde{\mathbf{v}} \rangle\right) \Big\vert,
\end{equation}
where we have defined $\tilde{\mathbf{v}} = \mathbf{v}/\sqrt{\langle \mathbf{v} \vert \mathbf{v} \rangle}$. As $\langle \tilde{\mathbf{v}} \vert \tilde{\mathbf{v}} \rangle = 1$, the argument of the logarithm is clearly bounded by the largest eigenvalue (by magnitude) $\lambda^{C}_{\rm Max}$ of $\hat{C}$. From $\vert c_{i} \vert  = \mathcal{O}(L^{\gamma}) \ \forall i$ we have $\vert \lambda^{C}_{\rm Max} \vert = \exp(\mathcal{O}(L^{\gamma}))$ and thus
\begin{equation}
    \Delta f = \Big\vert \frac{1}{L}\ln \left(\exp(\mathcal{O}(L^{\gamma})) \right) \Big\vert = \mathcal{O}(L^{\gamma - 1}),
\end{equation}
completing our proof of Lemma \ref{Lemma:SMLemma1} (the sign on the argument of the exponential $\exp(\mathcal{O}(L^{\gamma}))$ is not of concern as we are taking the absolute value of its logarithm).
\par Now that we have proven Lemma \ref{Lemma:SMLemma1} we will proceed to show that the largest magnitude eigenvalue of the difference matrix $\hat{H}(\mathcal{G}_{\rm ER}(p)) - \hat{H}(\mathcal{G}_{\rm Complete})$ scales as $\mathcal{O}(L^{1/2})$ with system size $L$. This will allow us to utilise Lemma \ref{Lemma:SMLemma1} with $\gamma = 1/2$ to prove Theorem \ref{Theorem:SMTH1}.
\par First, let us write $\hat{H}(\mathcal{G}_{\rm ER}(p))$ as $\hat{H}(\mathcal{G}) = L\sum_{\alpha}J_{\alpha} \cdot \hat{O}^{\alpha}(\mathcal{G}) + \hat{C}$ where $\hat{C}$ is a graph independent operator corresponding to the single-site terms and $\hat{O}^{\alpha}(\mathcal{G})$ is defined as follows.
\begin{equation}
    \hat{O}^{\alpha}(\mathcal{G}) = \frac{1}{N_{E}}\sum_{(v,v') \in E}\hat{s}^{\alpha}_{v}\hat{s}^{\alpha}_{v'}.
    \label{Eq:SMTwoBodySpinS}
\end{equation}
We now provide the following Lemma involving $\hat{O}^{\alpha}(\mathcal{G})$
\begin{lemma}
    Let $L$ be the number of vertices in $\mathcal{G}_{\rm ER}(p)$ and $\mathcal{G}_{\rm Complete}$. Let $\vert \psi \rangle$ be \textit{any} state such that $\langle \psi \vert \psi \rangle = 1$. Then $\big \vert \langle \psi \vert \hat{O}^{\alpha}(\mathcal{G}_{\rm ER}(p))  \vert \psi \rangle - \langle \psi \vert \hat{O}^{\alpha}(\mathcal{G}_{\rm Complete}) \vert \psi \rangle \big \vert = \mathcal{O}(L^{-1/2})$.
    \label{SMLemma:TwoBodyOp}
\end{lemma}
If we can prove this Lemma then, as $\hat{H}(\mathcal{G}_{\rm ER}(p)) - \hat{H}(\mathcal{G}_{\rm Complete}) = L\sum_{\alpha}J_{\alpha}(\hat{O}^{\alpha}(\mathcal{G}_{\rm ER}(p)) - \hat{O}^{\alpha}(\mathcal{G}_{\rm Complete}))$, the largest magnitude eigenvalue of the difference matrix $\hat{H}(\mathcal{G}_{\rm ER}(p)) - \hat{H}(\mathcal{G}_{\rm Complete})$ will be bounded as $\mathcal{O}(L^{1/2})$, letting us invoke Lemma \ref{Lemma:SMLemma1} to prove Theorem \ref{Theorem:SMTH1} as desired.  
\subsection*{Proof of Lemma \ref{SMLemma:TwoBodyOp} for Spin 1/2}
We set $\hat{s}^{\alpha}_{v} = \frac{1}{2}\hat{\sigma}^{\alpha}_{v}$, where $\hat{\sigma}^{\alpha}_{v}$ is the Pauli matrix for spin $\alpha = x,y$ or $z$ on vertex $V$. The dimension of the Hilbert space for a given $L$ is thus $2^{L}$.
In this context, the two-body operator $\hat{O}^{\alpha}(\mathcal{G})$ reads (we have dropped the factor of $s^{2}$ as it is inconsequential)
\begin{equation}
    \hat{O}^{\alpha}(\mathcal{G}) = \frac{1}{N_{E}}\sum_{(v,v') \in E}\hat{\sigma}^{\alpha}_{v}\hat{\sigma}^{\alpha}_{v'}.
    \label{Eq:SMTwoBody}
\end{equation}
\par In order to prove Lemma \ref{SMLemma:TwoBodyOp} we choose to work in the eigenbasis $\vert \sigma^{\alpha}_{1}, ..., \sigma^{\alpha}_{L} \rangle$ which diagonalises both $\hat{O}^{\alpha}(\mathcal{G}_{\rm ER}(p))$ and $\hat{O}^{\alpha}(\mathcal{G}_{\rm Complete})$, with $\sigma^{\alpha}_{v} = \pm 1$. The eigenvalues for $\hat{O}^{\alpha}(\mathcal{G}_{\rm Complete})$ are straightforward as we can define the `total spin' scalar $M_{\alpha} = \sum_{v}\sigma^{\alpha}_{v}$ to see that
\begin{align}
    \langle \sigma^{\alpha}_{1}, ..., \sigma^{\alpha}_{L} \vert \hat{O}^{\alpha}(\mathcal{G}_{\rm Complete}) \vert \sigma^{\alpha}_{1}, ..., \sigma^{\alpha}_{L} \rangle = \frac{1}{L(L-1)}\left((M_{\alpha})^{2} - L \right) = \left(\frac{M_{\alpha}}{L}\right)^{2} + \mathcal{O}(L^{-1}),
    \label{Eq:SMEigs}
\end{align}
where we have used the fact $\sigma^{\alpha}_{v}\sigma^{\alpha}_{v} \equiv 1$ and $N_{E} = \frac{1}{2}L(L-1)$ for the complete graph.
\par For $\hat{O}^{\alpha}(\mathcal{G}_{\rm ER}(p))$ we will utilise the notion of graph cuts to determine the corresponding eigenvalue of the state $\vert \sigma^{\alpha}_{1}, ..., \sigma^{\alpha}_{L} \rangle$. Specifically, this state can be considered a bi-partition of the underlying Erd\H{o}s-R\'enyi graph with the vertices for which $\sigma^{\alpha}_{v} = 1$ corresponding to one set ($A$) and the vertices where $\sigma^{\alpha}_{v} = -1$ being the other ($B$). The partition sizes are $(L + M_{\alpha})/2$ and $(L - M_{\alpha})/2$ respectively, with $M_{\alpha} = \sum_{v}\sigma^{\alpha}_{v}$.
\par We then have
\begin{equation}
    \langle  \sigma^{\alpha}_{1}, ..., \sigma^{\alpha}_{L} \vert \hat{O}^{\alpha}(\mathcal{G}_{\rm ER}(p))\vert \sigma^{\alpha}_{1}, ..., \sigma^{\alpha}_{L} \rangle = N_{E}(\mathcal{G}_{\rm ER}(p)) - 2N_{AB},
    \label{Eq:SMRHS}
\end{equation}
where $N_{E}(\mathcal{G}_{\rm ER}(p))$ is the number of edges associated with the given ER graph, $N_{AB}$ is the number of edges between the sets and is often referred to the `cut-size' of the partition. 
\par Because all of the sites and edges in an Erd\H{o}s-R\'enyi graph are independent a central limit theorem applies \cite{ErdosRenyiCuts1, ErdosRenyiCuts2} and the cut-size $N_{AB}$ for a randomly selected state $\vert \sigma^{\alpha}_{1}, ..., \sigma^{\alpha}_{L} \rangle$ will be a binomially distributed random variable with number of trials $n= (L^{2}-M_{\alpha}^{2})/4$ and probability $p$, i.e. $N_{AB} \sim B((L^{2}-M_{\alpha}^{2})/4, p)$.
The mean $\mu$ of this Binomial distribution is $p(L^{2}-M_{\alpha}^{2})/4$ and the standard deviation is $\sqrt{p(1-p)(L^{2}-M_{\alpha}^{2})}/2$. We wish to quantify the most extreme cuts which can occur with non-vanishing probability and thus place strict bounds on the deviation of $N_{AB}$ from its mean.
Such extremes can be found by applying a Chernoff bound \cite{ChernoffBound} on the tails of the Binomial distribution. Importantly, we then need to take a union bound over all cuts because we wish to ensure that this bound is tight for all $2^{L}$ states and thus valid for the whole spectrum of $\hat{O}^{\alpha}(\mathcal{G}_{\rm ER}(p))$. 
\par To make the application of these bounds explicit let us consider the case when $M_{\alpha} = 0$. Following the Chernoff bound we have the probability that a given cut size $N_{AB}$ exceeds the mean $\mu$ by a factor of $\delta$ is bounded via
\begin{equation}
    P(N_{AB} \geq (1+\delta)\mu) \leq e^{-\delta^{2}\mu/3},
\end{equation}
with $0 \leq \delta \leq 1$. We need, however, to ensure that this bound is tight for all $\binom{L}{L/2}$ cuts where $M_{\alpha} = 0$. Thus we can apply the union bound to get
\begin{equation}
    P'(N_{AB} \geq (1+\delta)\mu) \leq \binom{L}{L/2}e^{-\delta^{2}\mu/3},
\end{equation}
where $P'(N_{AB} \geq (1+\delta)\mu)$ is the probability that, of all the cuts with $M_{\alpha} = 0$, at least one of them deviates exceeds the mean by a factor of $\delta$. For large $L$ we can use a continuity approximation on the Binomial coefficient and reintroduce $\mu = p(L^{2}-M_{\alpha}^{2})/4$ to get
\begin{equation}
    P'(N_{AB} \geq (1+\delta)p(L^{2}-M_{\alpha}^{2})/4) \leq \sqrt{\frac{2}{\pi L}}e^{L\log_{e}(2) -p \delta^{2} L^{2}/12}.
    \label{Eq:SMChernoffResult}
\end{equation}
It is clear this probability vanishes unless $\delta \leq O(L^{-1/2})$. Therefore $N_{AB}$ can never exceed the mean by more than $O(L^{3/2})$ which vanishes in comparison to the mean. An identical result can be obtained for the lower end of the Binomial distribution. A similar result can also be obtained for any value of $M$ and we can apply the union bound over all possible $M$ values without weakening the result due to the strength of $e^{L}$ in comparison to a sum which only extends linearly in $L$.
\par Further details on bounding the spectrum of cuts in ER graphs can be found in Refs. \cite{ErdosRenyiCuts1, ErdosRenyiCuts2}. Formally, we can state that for absolutely \textit{any} state $\vert \sigma^{\alpha}_{1}, ..., \sigma^{\alpha}_{L} \rangle$
\begin{equation}
    \vert N_{AB} -\frac{1}{4}p (L^{2}-M_{\alpha}^{2}) \vert =\mathcal{O}(L^{3/2}).
    \label{Eq:SMEigenvalueBound}
\end{equation}
It therefore follows that $\lim_{L \rightarrow \infty}N_{AB}/L^{2} = \frac{1}{4}p (L^{2}-M_{\alpha}^{2})$. We also know that $N_{E}(\mathcal{G}_{\rm ER}(p)) = \frac{1}{2}pL^{2} + \mathcal{O}(L)$. Combining these results together we find that for absolutely \textit{any} state $\vert \sigma^{\alpha}_{1}, ..., \sigma^{\alpha}_{L} \rangle$
\begin{align}
   \langle \sigma^{\alpha}_{1}, ..., \sigma^{\alpha}_{L} \vert \hat{O}^{\alpha}(\mathcal{G}_{\rm ER}(p)) \vert \sigma^{\alpha}_{1}, ..., \sigma^{\alpha}_{L} \rangle = \frac{1}{N_{E}(\mathcal{G}_{\rm ER}(p))}(N_{E}(\mathcal{G}_{\rm ER}(p)) - 2N_{AB}) = \left(\frac{M_{\alpha}}{L}\right)^{2} + \mathcal{O}(L^{-1/2}).
    \label{Eq:SMExpecER}
\end{align}
Lemma \ref{SMLemma:TwoBodyOp} immediately follows from Eqs. (\ref{Eq:SMEigs}) and (\ref{Eq:SMExpecER}) -- the proof of Theorem \ref{Theorem:SMTH1} is thus complete for $s= 1/2$

\subsection*{Spin s}
We can extend our proof to all finite spin $s$. Let us work with the general spin $s$ definition of $\hat{O}^{\alpha}(\mathcal{G})$:
\begin{equation}
    \hat{O}^{\alpha}(\mathcal{G}) = \frac{1}{N_{E}}\sum_{(v,v') \in E}\hat{s}^{\alpha}_{v}\hat{s}^{\alpha}_{v'},
    \label{Eq:SMTwoBodySpinS}
\end{equation}
where $s^{\alpha}_{v}$ is the canonical spin $s$ operator on vertex $V$.
We work in the basis in which $\hat{O}^{\alpha}(\mathcal{G})$ is diagonal. The relevant basis states are  $\vert s^{\alpha}_{1}, ..., s^{\alpha}_{L} \rangle$ with $s^{\alpha}_{v}$ taking values in the set $\{-s, -s +1, ..., s\}$. 
\par We now define the scalar $M_{\alpha} = \sum_{v}s^{\alpha}_{v}$ and obtain
\begin{align}
    \langle s^{\alpha}_{1}, ..., s^{\alpha}_{L} \vert \hat{O}^{\alpha}(\mathcal{G}_{\rm Complete}) \vert s^{\alpha}_{1}, ..., s^{\alpha}_{L} \rangle = \frac{1}{L(L-1)}\left((M_{\alpha})^{2} - L \right) = \left(\frac{M_{\alpha}}{L}\right)^{2} + \mathcal{O}(L^{-1}),
    \label{Eq:SMEigsSpins}
\end{align}
in  direct analogy with the spin $1/2$ case. We can also diagonalise $\hat{O}^{\alpha}(\mathcal{G}_{\rm ER}(p))$ with the computational basis $\vert s^{\alpha}_{1}, ..., s^{\alpha}_{L} \rangle$. A given basis-state represents a partition of the vertices of the graph into $2s+1$ sets, with each set containing vertices which have the same value for $s^{\alpha}_{v}$. We therefore have
\begin{align}
    \langle s^{\alpha}_{1}, ..., s^{\alpha}_{L} \vert \hat{O}^{\alpha}(\mathcal{G}_{\rm Complete}) \vert s^{\alpha}_{1}, ..., s^{\alpha}_{L} \rangle =\frac{1}{N_{E}(\mathcal{G}_{\rm ER}(p))}\left(\sum_{i=-s}^{s}i^{2}N^{(i)}_{E} + \sum_{i=-s}^{s}\sum_{j=i+1}^{s}ijN^{(i,j)}_{E}\right),
    \label{Eq:SMNPartitions}
\end{align}
where $N_{E}^{(i)}$ is the number of edges present between all pairs of vertices with the same spin $i$, with $i \in [-s, -s +1, ..., s]$. Additionally, $N_{E}^{(i,j)}$ is the number of edges between pairs of vertices where one has spin $i$ and the other spin $j$. 
\par We now note that each term inside the summations in Eq. (\ref{Eq:SMNPartitions}) is independent of the other terms due to the independence of edges. We can then, just like for the spin $s=1/2$ case, invoke the Binomial distribution of edges, both within a set and between any two sets, and apply bounds on the tails of the distribution. From this it then follows that the integers $N_{E}^{(i)}$ and $N_{E}^{(i,j)}$ cannot deviate in any significant (i.e. quadratic in $L^{2}$) way from their expected values, which are $p(N_{i})^{2}/2$ and $pN_{i}N_{j}$ respectively --- with $N_{i}$ the number of vertices with spin $i$ in the state $\vert s^{\alpha}_{1}, ..., s^{\alpha}_{L} \rangle$. We also have $N_{E}(\mathcal{G}_{\rm ER}(p)) = \frac{1}{2}pL^{2} + \mathcal{O}(L)$. We can thus reduce Eq. (\ref{Eq:SMNPartitions}) to
\begin{equation}
\langle s^{\alpha}_{1}, ..., s^{\alpha}_{L} \vert \hat{O}^{\alpha}(\mathcal{G}_{\rm Complete}) \vert s^{\alpha}_{1}, ..., s^{\alpha}_{L} \rangle = \frac{2}{pL^{2}}\left(\frac{1}{2}\sum_{i=-s}^{s}i^{2}p(N_{i})^{2} + \sum_{i=-s}^{s}\sum_{j=i+1}^{s}ijpN_{i}N_{j}\right) + \mathcal{O}(L^{-1/2}).
\end{equation}
Observing that $M^{\alpha} = \sum_{v}s^{\alpha}_{v} = \sum_{i=-s}^{s}iN_{i}$ leads us to, for absolutely any basis state $\vert s^{\alpha}_{1}, ..., s^{\alpha}_{L} \rangle$,
\begin{equation}
        \langle s^{\alpha}_{1}, ..., s^{\alpha}_{L} \vert \hat{O}^{\alpha}(\mathcal{G}_{\rm ER}(p)) \vert s^{\alpha}_{1}, ..., s^{\alpha}_{L} \rangle = \left(\frac{M_{\alpha}}{L}\right)^{2} + \mathcal{O}(L^{-1/2}),
    \label{Eq:SMEigsSpinsV2}
\end{equation}
which, when combined with Eq. (\ref{Eq:SMEigsSpins}) gives us proof of Lemma \ref{SMLemma:TwoBodyOp}. The proof of Theorem \ref{Theorem:SMTH1} is thus complete.

\section*{Numerical Verification of the Proof}

\begin{figure}[!t]
    \centering
    \includegraphics[width =\textwidth]{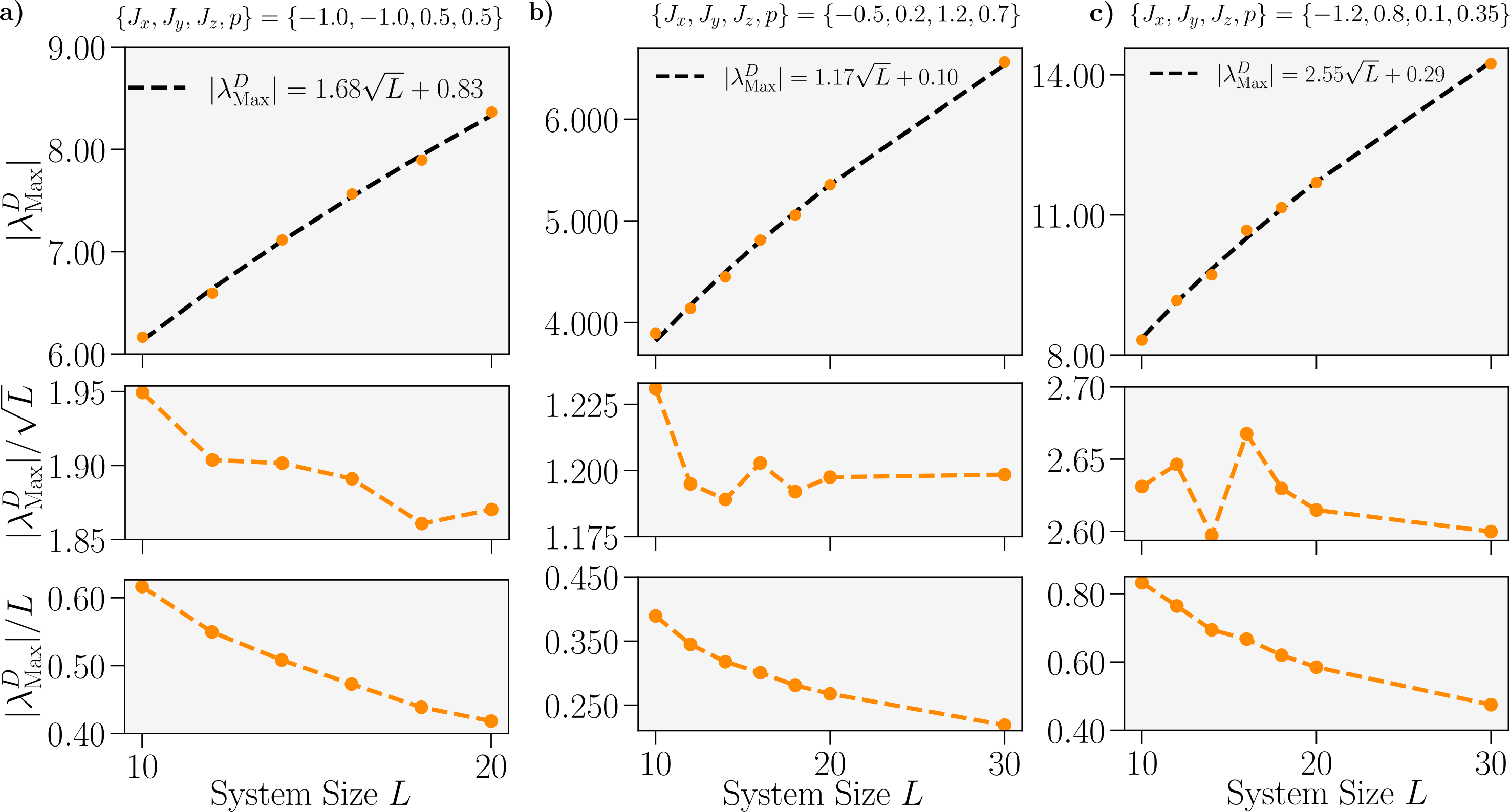}
    \captionsetup{width=\textwidth}
    \caption{Largest eigenvalue ( by absolute value) $\lambda^{D}_{\rm Max}$ of the difference matrix $\hat{D}(\mathcal{G}_{\rm ER}(p))$ -- see Eq. (\ref{Eq:SMDiffMat}) -- versus system size. a-c) Corresponds to several different sets of the parameters $\{J_{x}, J_{y}, J_{z}, p\}$.  Top plots provide the direct scaling of $\lambda^{D}_{\rm Max}$ with $L$ whilst the middle and bottom give the scaling of $\lambda^{D}_{\rm Max}/\sqrt{L}$ and $\lambda^{D}_{\rm Max}/L$ with $L$ respectively. Dashed black line in the top plots represents a fit to the corresponding data -- with the explicit equation given in the legend. Each data point corresponds to an average over $n=100$ draws of the ER graph from its ensemble. System sizes for $L \leq 16$ were obtained by Exact Diagonalisation whilst $L > 16$ were obtained using DMRG on $\pm \hat{D}(\mathcal{G}_{\rm ER}(p))$ with a bond-dimension $\chi = 10L$.}
    \label{Fig:SMFM1}
\end{figure}
\par We now provide numerical evidence to support Theorem \ref{Theorem:SMTH1}. In order to do so we consider the following matrix
\begin{equation}
    \hat{D}(\mathcal{G}_{\rm ER}(p)) =  \hat{H}(\mathcal{G}_{\rm Complete}) - \hat{H}(\mathcal{G}_{\rm ER}(p)) = \sum_{\alpha = x,y,z}J_{\alpha}\sum_{\substack{v, v' = 1 \\ v > v'}}^{L}\left(\frac{2}{L-1} - \frac{p_{v,v'}L}{N_{E}(\mathcal{G}_{\rm ER}(p))} \right)\hat{\sigma}^{\alpha}_{v}\hat{\sigma}^{\alpha}_{v'},
    \label{Eq:SMDiffMat}
\end{equation}
which is the difference between $H(\mathcal{G})$ on a given instance $\mathcal{G}_{\rm ER}(p)$ of the ER graph and $H(\mathcal{G})$ on the complete graph. Here, $p_{v,v'}$ is a random variable which is $1$ with probability $p$ and $0$ otherwise. The quantity $N_{E}(\mathcal{G}_{\rm ER}(p)) = \sum_{v > v'}p_{v,v'}$ is the number of edges of the given ER graph. 
\par To support Theorem \ref{Theorem:SMTH1} we numerically determine the scaling on the largest absolute eigenvalue $\lambda^{D}_{\rm Max}$ of $\hat{D}(\mathcal{G}_{\rm ER}(p))$ as a function of $L$ -- for several different choices of the parameter set $\{J_{x}, J_{y}, J_{z}, p\}$. In Supplementary Fig. \ref{Fig:SMFM1} we plot this scaling and observe a clear linear dependence of $\lambda^{D}_{\rm Max}$ on $\sqrt{L}$ -- with $\lambda^{D}_{\rm Max}/\sqrt{L}$ staying fairly constant with $L$ whilst $\lambda^{D}_{\rm Max}/L$ decays. It is this scaling on $\lambda^{D}_{\rm Max}$  which we have mathematically proven and allows us to invoke Lemma \ref{Lemma:SMLemma1} to prove Theorem \ref{Theorem:SMTH1}.

\section*{XXZ Hamiltonian on the Complete Graph}
We consider the XXZ spin $s = 1/2$ Hamiltonian on the complete graph, dropping the factor of $s^{2}$ on the two-body operators as it is irrelevant:
\begin{equation}
        \hat{H}_{\rm XXZ}(\mathcal{G}_{\rm Complete}) = \frac{2}{L-1}\sum_{\substack{v, v' =1 \\ v > v'}}^{L}-J(\hat{\sigma}_{v}^{x}\hat{\sigma}^{x}_{v'} + \hat{\sigma}_{v}^{y}\hat{\sigma}^{y}_{v'}) + \Delta \hat{\sigma}^{z}_{v}\hat{\sigma}^{z}_{v'}.
        \label{Eq:SMHXXZComplete}
\end{equation}
\par We focus on the case where $J$ and $\Delta$ are positive, finite, real numbers. In the limit $L \rightarrow \infty$ the free energy density for this Hamiltonian is also (by Theorem \ref{Theorem:SMTH1}) that for the  Hamiltonian $\hat{H}_{\rm XXZ}(\mathcal{G}_{\rm ER}(p))$ with finite edge probability $p$. We can thus focus exclusively on $\hat{H}_{\rm XXZ}(\mathcal{G}_{\rm Complete})$ to quantify the equilibrium properties of both cases as $L \rightarrow \infty$.
We can diagonalise this Hamiltionian by introducing the global operators $\hat{S}^{\alpha} = \sum_{v}\hat{\sigma}^{\alpha}_{v}$, the Casimir operator $\hat{S}^{2} = (\hat{S}^{x})^{2} +(\hat{S}^{y})^{2} + (\hat{S}^{z})^{2}$ and the complete total-spin basis $\vert S, M \rangle$ with $\hat{S}^{2}\vert S,M \rangle = 4S(S+1)\vert S,M \rangle$ and $\hat{S}^{z}\vert S,M \rangle = M \vert S,M \rangle$. The integer $S$ can take values in the range $\{M/2, M/2 + 1, ..., L/2\}$ whilst $M$ can take values in the range $\{-L, - L +2, ..., L\}$ \cite{LMGModel1} --- assuming $L$ is even. Following a small amount of algebra we have
\begin{align}
        &\hat{H}_{\rm XXZ}(\mathcal{G}_{\rm Complete})\vert S,M \rangle = \lambda_{S,M}\vert S,M \rangle \notag \\
    &\lambda_{S,M} = -\frac{J}{L-1}(4S(S+1)-M^{2}) + \frac{2JL}{L-1} + \frac{\Delta}{L-1}(M^{2}-L), \notag \\
    &D_{S,M} =
        \binom{L-1}{L/2 -S - \delta_{S, 0}} - \binom{L-1}{L/2 -S - 1 - \delta_{S, 0}},
    \label{Eq:SMAnalyticalEigs}
\end{align}
where $\delta_{i,j}$ is the Kronecker Delta function and $D_{S,M}$ is the degeneracy of a given $\lambda_{S,M}$ eigenvalue in the full $2^{L}$ dimensional Hilbert space. These degeneracies were found via the Clebsch-Gordan coefficients for an SU(2) symmetry in a space formed from an $L$-fold tensor product of spin 1/2s \cite{PhysRevB.103.035146}.  
\par Taking the thermodynamic limit we find the spectrum of $\hat{H}_{\rm XXZ}(\mathcal{G}_{\rm Complete})$ will thus be dominated by states of $0$ energy density as the degeneracy of the $\ket{S,M}$ states is largest when $S$ and $M$ do not grow proportionally with $L$. In Supplementary Fig. \ref{Fig:SMFM2} we plot a histogram of this spectrum for a few choice system sizes. The probability density narrows around $0$ meaning that the system is dominated by states with $~0$ energy density.
\par The ground states meanwhile are far from these `typical' states and consists of any of the states $\vert L/2, M \rangle$ where $M$ is finite. More explicitly such states can be written as
\begin{equation}
    \vert \psi \rangle_{\rm GS} = \vert L/2,M \rangle \propto \sum_{\sigma_{1} + \sigma_{2} + ... \sigma_{L} = M}\vert \sigma_{1}, \sigma_{2}, ..., \sigma_{L} \rangle,
\end{equation}
i.e. they are an equal superposition of all basis states with $M$ total magnetisation.

\begin{figure}[!t]
    \centering
    \includegraphics[width =\textwidth]{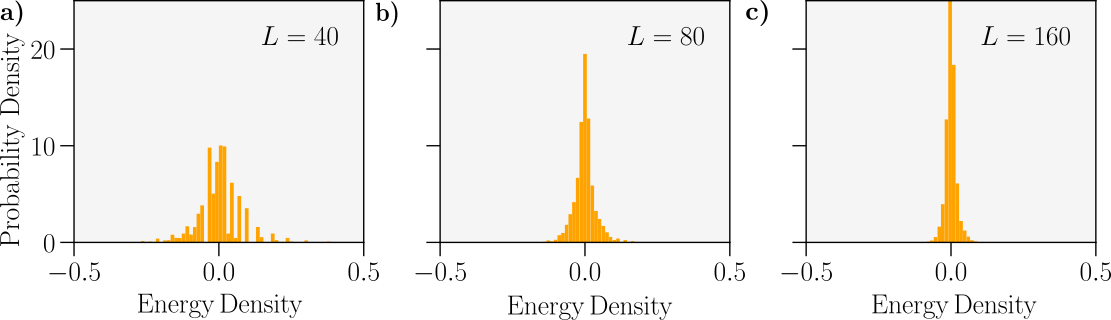}
    \captionsetup{width=\textwidth}
    \caption{Probability Density $p(E)$ for the eigenenergies of $\hat{H}_{\rm XXZ}(\mathcal{G}_{\rm Comp})$ -- see Eq. (\ref{Eq:SMHXXZComplete}) -- with $\Delta = 1.5$ and for various system sizes $L$. The quantity $p(E)dE$ is the probability of a randomly chosen eigenstate having an energy density in the interval $[p(E), p(E + dE)]$. The spectrum was calculated exactly via Eq. (\ref{Eq:SMAnalyticalEigs}) and as $L \rightarrow \infty$ tends towards a delta function which also represents the spectrum of $\hat{H}_{\rm XXZ}(\mathcal{G}_{\rm ER}(p))$ for any finite non-zero $p$.}
    \label{Fig:SMFM2}
\end{figure}

\section*{Spin Hamiltonian on an Arbitrary graph with a Non-Trivial Cut}
\subsection*{Proof of Reduction of Free Energy Density to That For Two Large Spins}
Here we consider $\hat{H}(\mathcal{G}(\lambda, p_{1}, p_{2}))$ where $\hat{H}$ is defined in Eq. (\ref{Eq:SMGeneralHamiltonian}) and $\mathcal{G}(\lambda, p_{1},p_{2})$ is an instance of the random graph with a non-trivial cut and $L$ vertices --- as defined in the Methods section of the main text. We introduce the two sets of vertices $A = \{1, ..., \lambda L\}$ and $B = \{\lambda L + 1, ..., L\}$ and define the following effective Hamiltonian from the parameters $\lambda, p_{1}$  and $p_{2}$
\begin{equation}
    \hat{H}_{\rm Eff}(\mathcal{G}(\lambda, p_{1}, p_{2})) = \frac{1}{N}\sum_{\alpha = x,y,z}J_{\alpha}\left( p_{1}(\hat{S}^{\alpha}_{A})^{2} + p_{1}(\hat{S}^{\alpha}_{B})^{2} + 2p_{2}\hat{S}^{\alpha}_{A}\hat{S}^{\alpha}_{B})   \right) + \frac{1}{N}\sum_{\alpha = x,y,z}w_{\alpha}(\hat{S}^{\alpha}_{A} + \hat{S}^{\alpha}_{B}),
\end{equation}
where $\hat{S}^{\alpha}_{A} = \sum_{v \in A}\hat{s}^{\alpha}_{v}$, $\hat{S}^{\alpha}_{B} = \sum_{v \in B}\hat{s}^{\alpha}_{v}$ and $N = \frac{1}{2}L\left(\lambda^{2}p_{1} + (\lambda - 1)^{2}p_{1} + 2\lambda (\lambda - 1)p_{2}\right)$.
\par With these definitions we prove the following theorem
\begin{theorem}
    For a given set of values for the microscopic parameters $\{J_{x}, J_{y}, J_{z}, w_{x}, w_{y}, w_{z}\}$ then
    \begin{equation}
        \lim_{L \rightarrow \infty}f(\hat{H}(\mathcal{G}(\lambda, p_{1}, p_{2}))) = \lim_{L \rightarrow \infty}f(\hat{H}_{\rm Eff}(\mathcal{G}(\lambda, p_{1}, p_{2}))),
    \end{equation}
    with $\vert f(\hat{H}(\mathcal{G}(\lambda, p_{1}, p_{2}))) - f(\hat{H}_{\rm Eff}(\mathcal{G}(\lambda, p_{1}, p_{2})))\vert = \mathcal{O}(L^{-1/2})$ for finite $L$.
    \label{Theorem:SMTheorem2}
\end{theorem}

In order to prove this theorem we focus on the two-body graph-dependent operators contained within $\hat{H}(\mathcal{G}(\lambda, p_{1}, p_{2}))$. These can be written as
\begin{equation}
        \hat{O}(\mathcal{G}(\lambda, p_{1},p_{2})) = \frac{1}{N_{E}(\mathcal{G}(\lambda, p_{1},p_{2}))}\sum_{\substack{v,v' =1 \\ v > v'}}^{L}\tilde{p}_{v,v'}\hat{s}^{\alpha}_{v}\hat{s}^{\alpha}_{v'},
        \label{Eq:SMNonTrivialOp}
\end{equation}
where $\tilde{p}_{v,v'}$ is a random variable which takes value $1$ with probability $p_{1}$ if the vertices $v$ and $v'$ belong to the same set of sites, takes value $1$ with probability $p_{2}$ if they belong to different sets, and takes value $0$ otherwise. The quantity $N_{E}(\mathcal{G}(\lambda, p_{1},p_{2}))$ is the number of edges on the instance of $\mathcal{G}(\lambda, p_{1}, p_{2})$.
\par Now we will prove the following Lemma
\begin{lemma}
    Let $\vert \psi \rangle$ be \textit{any} state such that $\langle \psi \vert \psi \rangle = 1$. Then the following holds
    \begin{equation}
    \langle \psi \vert \hat{O}(\mathcal{G}(\lambda, p_{1},p_{2})) \vert \psi \rangle =\left \langle \psi \left \vert \frac{1}{2N_{E}(\mathcal{G}(\lambda, p_{1},p_{2}))}\bigg(p_{1}\Big((\hat{S}^{\alpha}_{A})^{2} + (\hat{S}^{\alpha}_{B})^{2}\Big) + 2p_{2}\hat{S}^{\alpha}_{A}\hat{S}^{\alpha}_{B}\bigg) \right \vert \psi \right \rangle + \mathcal{O}(L^{-1/2}).
    \label{Eq:SMNonTrivialCutExpec}
    \end{equation}
    \label{SMLemma:Lemma5}
\end{lemma}
This Lemma will enable us to prove Theorem \ref{Theorem:SMTheorem2}. In order to prove Lemma \ref{SMLemma:Lemma5} observe that we can re-write Eq. (\ref{Eq:SMNonTrivialOp}) as
\begin{align}
        \hat{O}(\mathcal{G}(\lambda, p_{1},p_{2})) = \frac{1}{N_{E}(\mathcal{G}(\lambda, p_{1},p_{2}))}\left(\sum_{\substack{v,v' =1 \\ v > v'}}^{\lambda L}p^{1}_{v,v'}\hat{s}^{\alpha}_{v}\hat{s}^{\alpha}_{v'} + \sum_{\substack{v,v' =\lambda L  +1 \\ v > v'}}^{ L}p^{1}_{v,v'}\hat{s}^{\alpha}_{v}\hat{s}^{\alpha}_{v'} + \sum_{v = 1}^{\lambda L}\sum_{v'=\lambda L + 1}^{L}p^{2}_{v,v'}\hat{s}^{\alpha}_{v}\hat{s}^{\alpha}_{v'} \right),
        \label{Eq:SMEq20}
\end{align}
where $p^{1}_{v,v'}$ and $p^{2}_{v,v'}$  are random values which take value $1$ with probabilities $p_{1}$ and $p_{2}$ respectively and take value $0$ otherwise. The first two summations in the above expression are just two-body operators over an Erd\H{o}s-R\'enyi graph of size $\lambda L$ and $(1-\lambda) L$. We know from Lemma \ref{SMLemma:TwoBodyOp} we can replace their expectation with those with their collective counterparts with corrections of order $\mathcal{O}(L^{-1/2})$. In the final summation we can also, up to corrections of order $\mathcal{O}(L^{-1/2})$, replace $p^{2}_{v,v'}$ with $p_{2}$ and move it outside the sum. This again follows from identifying the normal distribution associated with a given eigenvalue of the operator and applying a Chernoff and Union bounds to find the most extreme possibilities. From this it follows that Lemma \ref{SMLemma:Lemma5} is true.
\par It is then clear from Weyl's inequality that the largest magnitude eigenvalue of the difference matrix $D = \hat{H}(\mathcal{G}(\lambda, p_{1},p_{2})) - \hat{H}_{\rm Eff}(\mathcal{G}(\lambda, p_{1}, p_{2}))$ is bounded as $\mathcal{O}(L^{1/2})$. We can now use $N_{E}(\mathcal{G}(\lambda, p_{1},p_{2})) = \frac{1}{2}L^{2}\left(\lambda^{2}p_{1} + (\lambda - 1)^{2}p_{1} + 2\lambda (\lambda - 1)p_{2}\right) + \mathcal{O}(L)$ and invoke Lemma \ref{Lemma:SMLemma1} to arrive at the proof of \ref{Theorem:SMTheorem2}.
\subsection*{Critical Point of the XXZ Limit of $\hat{H}(\mathcal{G}(\lambda, p_{1},p_{2}))$}
\par We now consider the XXZ limit of $\hat{H}(\mathcal{G}(\lambda, p_{1},p_{2}))$ and derive the location of the critical point. Theorem \ref{Theorem:SMTheorem2} tells us, in the thermodynamic limit, the equilibrium properties of this system are equivalent to those derived from the simpler Hamiltonian
\begin{align}
    &\hat{H}_{\rm Eff, XXZ}(\mathcal{G}(\lambda, p_{1},p_{2}))  = \notag \\ &\frac{1}{N}\bigg(-Jp_{1}\left(\sum_{\alpha = x,y}(S^{\alpha}_{A})^{2} + (S^{\alpha}_{B})^{2}\right) + \Delta p_{1} \left((\hat{S}^{z}_{A})^{2}) + (\hat{S}^{z}_{B})^{2}\right) - 2Jp_{2}\left(\sum_{\alpha = x,y}\hat{S}^{\alpha}_{A}\hat{S}^{\alpha}_{B}\right) + \Delta p_{2}\hat{S}^{z}_{A}\hat{S}^{z}_{B})\bigg),
\end{align}
for two large spins.
\par We can minimise this energy with a classical solution where the collective spins $A$ and $B$ are described by vectors $(s^{x}_{A}, s^{y}_{A}, s^{z}_{A})$ and $(s^{x}_{B}, s^{y}_{B}, s^{z}_{B})$ with magnitudes $s\lambda L$ and $s(1-\lambda)L$ respectively. The energy is then minimised by polarising both spins $A$ and $B$ in either the $+y$ or $+x$ directions for $\Delta < \Delta_{c}$ and polarising spins $A$ and $B$ in opposite directions along the $z$ axis for $\Delta > \Delta_{c}$. Comparing the energies of these states yields the critical point
\begin{equation}
    \Delta_{c} = \frac{p_{1}(\lambda^{2} + (1-\lambda)^{2}) + p_{2}\lambda(1-\lambda)}{2p_{2}\lambda(1-\lambda) - p_{1}(\lambda^{2} + (1-\lambda)^{2})}J,
\end{equation}
which is valid for any finite $s$.
Substituting the values $\lambda = 1/2$, $p_{1} = 1/2$, $p_{2} = 1$ and $J = 1$ which we use for Fig. 3 of the main text yields $\Delta_{c} = 3$ -- in agreement with our numerical data.

\begin{figure}[!t]
    \centering
    \includegraphics[width =\textwidth]{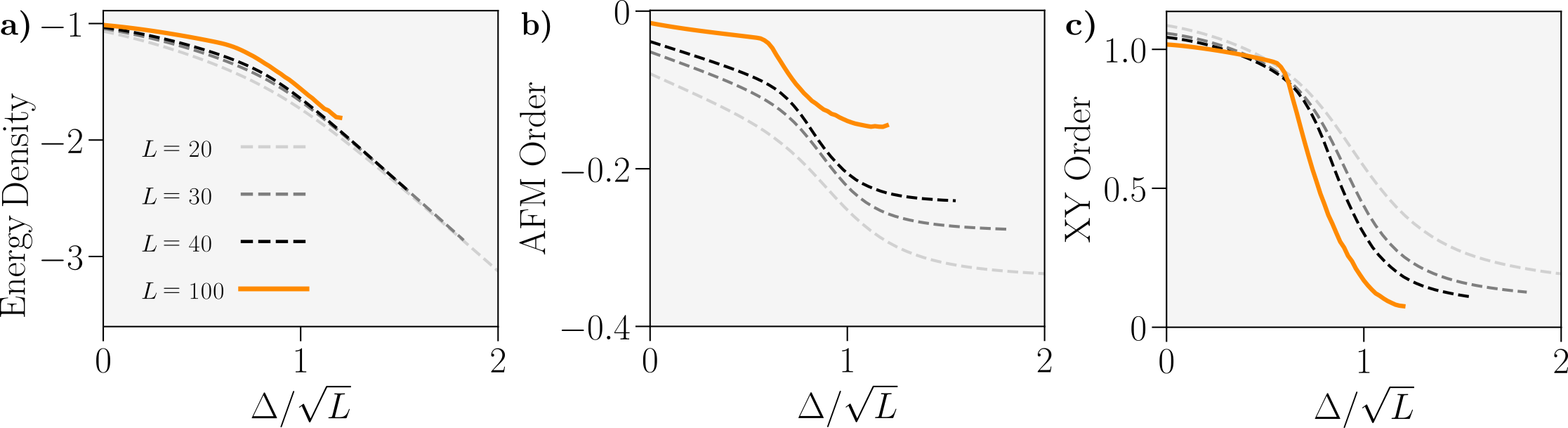}
    \captionsetup{width=\textwidth}
    \caption{Reproduced from the main-text but on a re-scaled $x$-axis. Properties of the ground-state of the XXZ Hamiltonian on the Erd\H{o}s-R\'enyi (ER) graph with $p = 0.5$.
    \textbf{a-c)} Energy density, anti-ferromagnetic order  $C_{\rm AFM}$ and XY order $C_{\rm XY}$  versus $\Delta / \sqrt{L}$, with the system size $L$ provided in the legend.}
    \label{Fig:SMF0}
\end{figure}

\section*{Erd\H{o}s-R\'enyi Plots with Re-scaled Axes}
Here, in Supplementary Fig. \ref{Fig:SMF0}, we re-plot the results of Fig. 2 in the main-text --- which correspond to the ground-state properties of the spin $s=1/2$ XXZ model on the Erd\H{o}s-R\'enyi graph --- but with the $x$-axis rescaled by $\sqrt{L}$.
We observe that the re-scaling of $\sqrt{L}$ is sufficient to prevent the drift of the critical point seen in the main text. Whilst accurately estimating this critical point is beyond the scope of this work the plots suggests the convergence of critical point to a value in the region $0 < \Delta/\sqrt{L} < 1$. 

\section*{Numerical Details for Matrix Product State Calculations}
Here we provide details for the Matrix Product State (MPS) calculations we performed for the XXZ model on various graphs. We utilised the Tensor Network Python (TeNPy) library \cite{TENPY1} for our simulations, encoding the initial guess for ground-state as an open-boundary MPS and the Hamiltonian as a long-range Matrix Product Operator (MPO) \cite{MPOLongRange1}. Using the Density Matix Renormalisation Group (DMRG) \cite{DMRG1} algorithm with single site updates and a mixer enabled we then iteratively improved on our initial guess until convergence was achieved for the specified bond-dimension $\chi$.
\par We used a direct mapping from the $v = 1 .. L$ sites of the graph to the sites of the MPS $s = 1 ...L$. For the graph $\mathcal{G}(\lambda = 1/2, p_{1}=1,p_{2}=1/2)$ we set the first $L/2$ sites of the graph to correspond to one partition and the other $L/2$ to the other (see adjacency matrix in Fig 3 in the main text), no ordering was done beyond this. For the maximally irregular graph we set the site with the highest degree to the central site ($V= L/2$) and have the distance of other sites to the central site increase with decreasing degree (leaving the sites with degrees $1$ and $2$ as $V = 1$ and $V = L$ respectively). This can be observed in the adjacency matrix in Fig. 4 of the main text. For all other graphs no attempts were made to order the sites of the graph in any specific way and thus the same is true for the sites of the MPS. 

\par \textit{Truncation Errors} - For a given DMRG run we calculate the total truncation error, i.e. the sum of the squares of all singular values discarded during the final sweep of the routine and in Supplementary Fig. \ref{Fig:SMF1} we plot these for the simulations performed in the main text. For random graphs where the ground state is found on multiple instances of the graph and averaging is performed, we plot the average over these instances.
\par The truncation error is never greater than $1.2 \times 10^{-4}$. Combining this with the smoothness of the observables as a function of $\Delta$ suggests our results are reasonably accurate and well-converged. The values of $\chi$ used for the highest system sizes represent the maximum possible with the computational resources available to us (for $L = 100$ and $\chi = 1600$ we required around 20GB of RAM for each simulation, setting a limit on the available $\chi$).
\par Notably, we find for the results on the random graphs the truncation error is improved for values of $L \sim 100$ compared to $L \sim 40$ despite using a similar, or even smaller value of $\chi$. We understand this due to finite-size effects (which can be particularly strong due to the statistical fluctuations associated with drawing from a random ensemble) being less significant for the higher value of $L$. At such higher values of $L$ the results more accurately obey the scaling of the large $L$ solution which, for the dense graphs we study, typically has a bi-partite entanglement entropy that scales as ${\rm log}(L)$ and makes the ground-state accessible with a bond-dimension proportional to the system size. We note that such a scaling has previously been observed in the Lipkin-Meshov-Glick model \cite{LMGModel1} (an all-to-all XY model with a transverse field) and here we observe it for a range of dense graphs.
\par \textit{Energy Convergence} - In Supplementary Fig. \ref{Fig:SMF1} we also plot examples of the convergence of the ground-state energy obtained versus DMRG sweep number for the range of graphs we consider in the paper and the largest system sizes used. For each graph, we select values of $\Delta$ where the truncation errors are highest and show the convergence for values of $\chi$ up to that used in our results. All plots show a clear convergence in the ground-state energy with increasing $\chi$ and indicate the value of $\chi$ used in our work is reasonable, with the energy obtained via DMRG, changing by at most $0.729 \%$ (and generally by over an order of magnitude less) when we double the bond-dimension to reach the value used in our results.

\begin{figure}[!t]
    \centering
    \includegraphics[width =\textwidth]{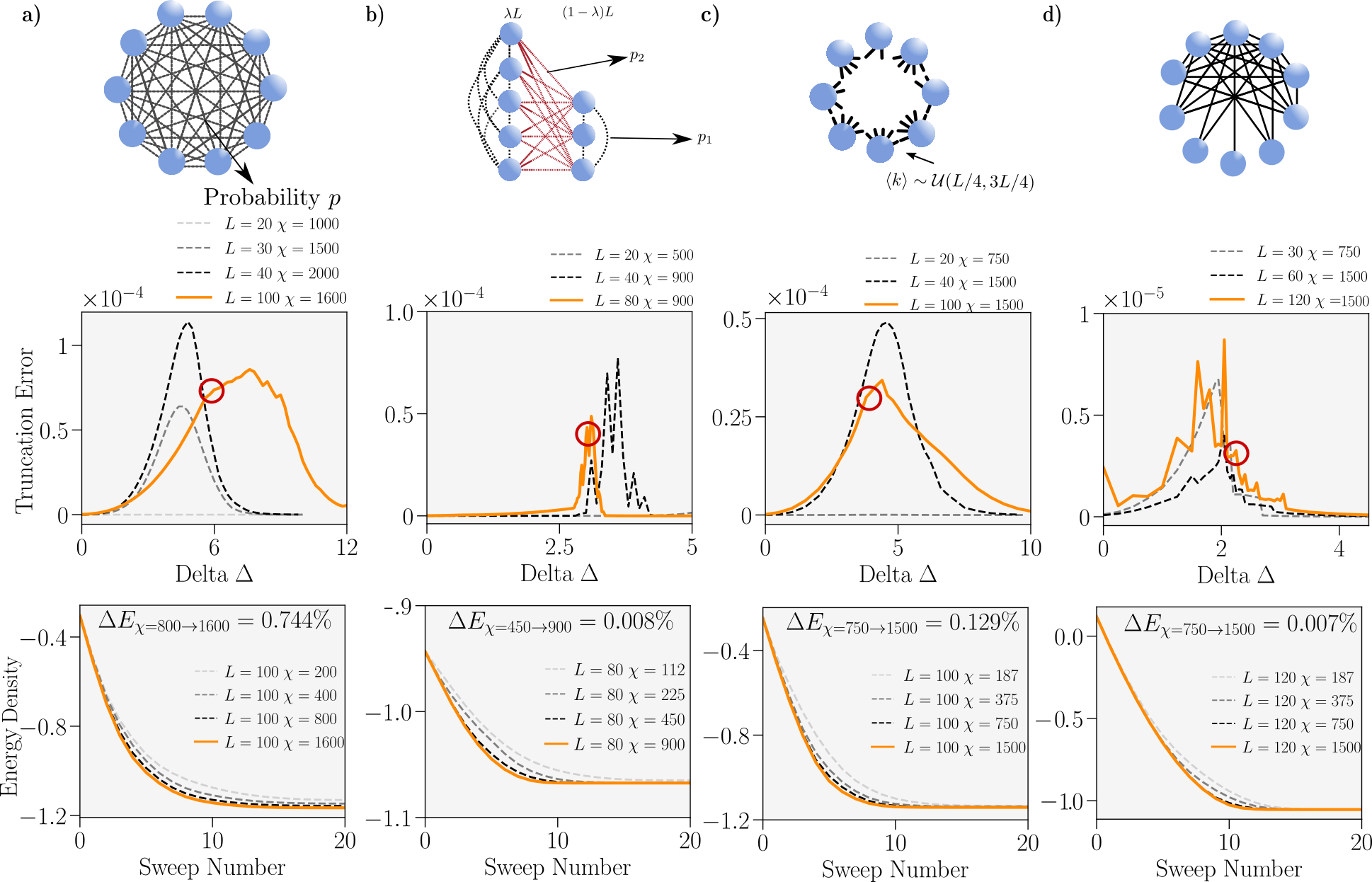}
    \captionsetup{width=\textwidth}
    \caption{Truncation error (total sum of the square of the singular values discarded during the DMRG routine) and energy versus sweep number for the ground-state calculations performed and whose results are provided in the main text.\textbf{a-d)} Graphs from Figures 2,3,4 and 5 of the main text respectively with graph schematics provided at the top. Vertically aligned plots correspond to the same graph, which is illustrated at the top along with a legend for the system sizes and bond-dimension used. Upper plots) Truncation error for a range of $\Delta$ values. For the graphs drawn from a random ensemble the truncation error is averaged over the same number of draws as in the main text. Lower Plots) Energy Density versus sweep number for the DMRG routine for the highest value of $L$ used and for the $\Delta$ values $6.0, 2.9, 4.5$ and $2.3$ respectively (marked in red in the plot above). The legend in the plot indicates the bond-dimensions used and we also annotate the percentage change in energy density $\Delta E$ when doubling the bond-dimension to reach the value used in our work. For the graphs drawn from a random ensemble the plots correspond to a single instance which was selected in a completely unbiased manner. }
    \label{Fig:SMF1}
\end{figure}

\begin{figure}[!t]
    \centering
    \includegraphics[width =\textwidth]{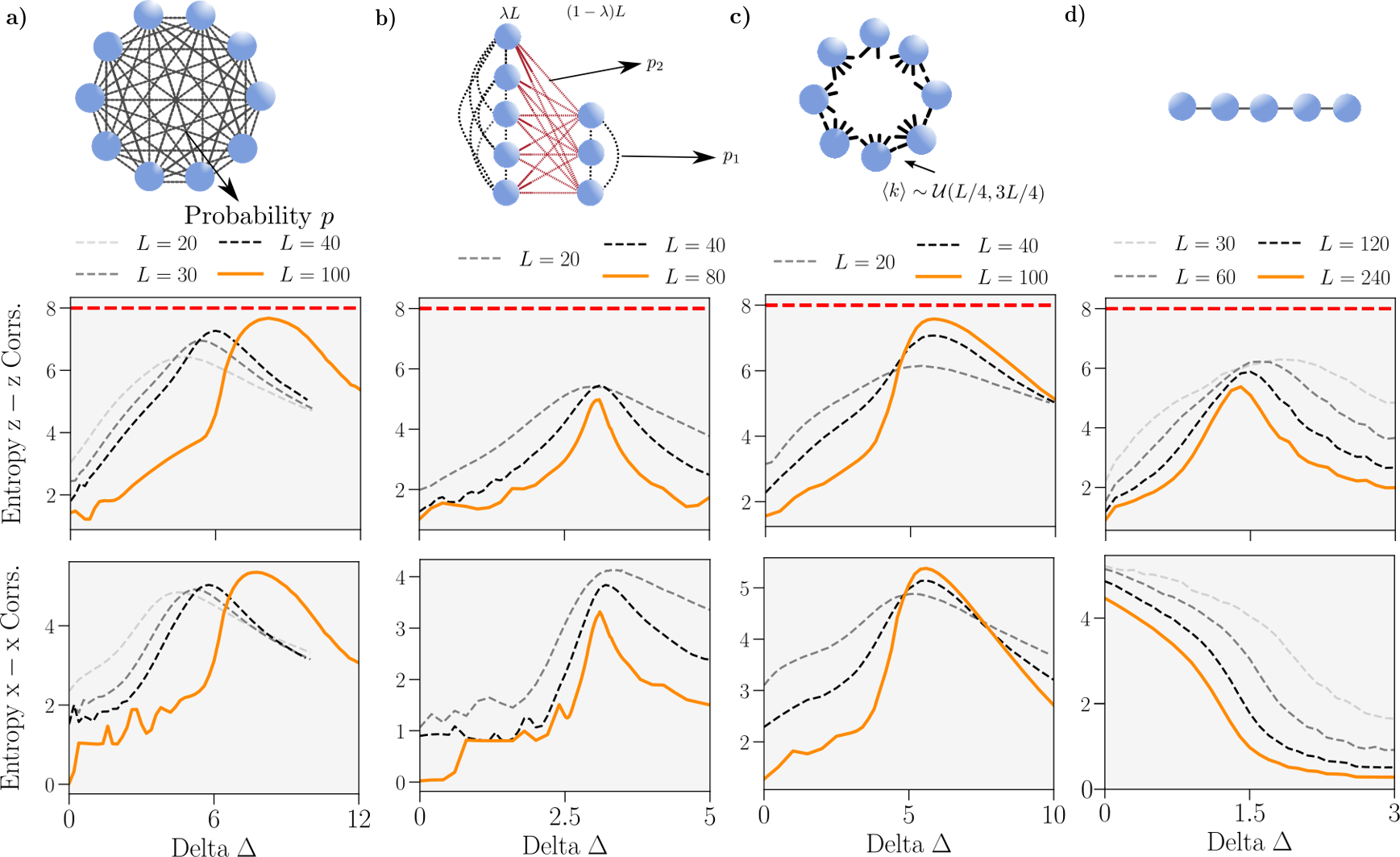}
    \captionsetup{width=\textwidth}
    \caption{Shannon Entropy --- see Eq. (\ref{Eq:SMShannonEntropy}) --- of the off-diagonal correlations in the ground-state of the XXZ model versus $\Delta$ for various different graphs and system sizes. We use $n = 256$ bins for the off-diagonal correlations.  \textbf{a-d)} Graphs used in Figures 2, 3, 4 of the main text and the 1D Chain respectively. Upper plots are the entropy $H(\langle \hat{\sigma}^{z}_{v}\hat{\sigma}^{z}_{v'}\rangle)$ whilst the lower are $H(\langle \hat{\sigma}^{x}_{v}\hat{\sigma}^{x}_{v'}\rangle)$. Red-dotted line indicates the maximum possible Shannon Entropy for the $n = 256$ bins used. Above the plot we provide illustrations of the respective graphs and colour code for the system sizes. We adopt the same averaging over random instances of the graphs as in the main text and the same bond dimensions as specified in Supplementary Fig. \ref{Fig:SMF1}. No averaging is required for the chain as it is unique for a given $L$. The bond dimension used for the chain is $\chi = 500$ for all system sizes.}
    \label{Fig:SMF2}
\end{figure}
\par \textit{Calculation of the Shannon Entropy for various graphs} - In the main text we also calculated, for the maximally irregular graph, the Shannon entropy which is defined as 
\begin{equation}
    H(\langle \hat{\sigma}^{\alpha}_{v}\hat{\sigma}^{\alpha}_{v'}\rangle) = \sum_{i = 0}^{n-1}p_{i}\log_{2}(p_{i}),
    \label{Eq:SMShannonEntropy}
\end{equation}
where $p_{i}$ is the fraction of elements of the off-diagonal $L \times L$ matrix $\langle \hat{\sigma}^{\alpha}_{v} \hat{\sigma}^{\alpha}_{v'}\rangle$ whose values are in the range $[-1 + 2i/n, -1 + 2(i+1)/n]$. Here, in Supplementary Fig. \ref{Fig:SMF2}, we plot this quantity (setting $n = 256$) for all the graphs considered in the main text (aside from the maximally irregular which is already plotted in Fig. 5 of the main text) and the 1D chain. We compute the entropy along both the $\alpha =x$ and $\alpha = z$ spin axes.
\par In the Erd\H{o}s-R\'enyi graph we find the entropy in both spin axes is maximised at the critical point and this maximum is not diminishing. We know, however, that the critical point diverges with $L$ and the ground state is always the Dicke-state $\ket{L/2, M}$ for finite $\Delta$ as $L \rightarrow \infty$. The Shannon entropies $H(\langle \hat{\sigma}^{x}_{v}\hat{\sigma}^{x}_{v'}\rangle)$ and $H(\langle \hat{\sigma}^{y}_{v}\hat{\sigma}^{y}_{v'}\rangle)$ are both zero for this state as all off-diagonal correlations take the same value. Both entropies will thus be $0$ on the ER graph for all finite $\Delta$ as $L \rightarrow \infty$. For the random graph with a non-trivial cut, i.e. $\mathcal{G}(\lambda, p_{1},p_{2})$ we find the entropy in both axes is diminishing with system sizes and far from it's maximal value. Moreover, we know that for $L \rightarrow \infty$  the system's ground state is the Dicke state for $\Delta < \Delta_{c}$ and an anti-ferromagnet for $\Delta > \Delta_{c}$. The entropy of the $x-x$ correlations will thus always be $0$ and the entropy of the $z-z$ correlations will be $0$ for $\Delta < \Delta_{c}$ and $\lambda \log_{2}(\lambda) +(1-\lambda) \log_{2}(1-\lambda)$ for $\Delta > \Delta_{c}$, which has an upper bound of $1$. For the irregular dense graph with $k \sim \mathcal{U}(L/4, 3L/4)$ we find qualtitatively the same features as the maximally irregular graph in the main text: the entropy in both spin axes is maximal at the critical point and not diminishing with system size. For the 1D chain we find the entropy in both spin axes is diminishing with system size. Moreover, because there is no long-range $x-x$ order in the ground-state in the thermodynamic limit we know the $x-x$ entropy must vanish as measuring $\langle \hat{\sigma}^{x}_{v}\hat{\sigma}^{x}_{v'}\rangle$ for a random pair of sites as $L \rightarrow \infty$ will yield $0$ with probability tending to $1$.

\section*{Transverse Field Ising Model}
\par To supplement our numerical results for the XXZ limit of the Hamiltonian in Eq. (\ref{Eq:SMGeneralHamiltonian}) we have also performed numerical calculations for the transverse field Ising (TFI) limit of Eq. (\ref{Eq:SMGeneralHamiltonian}). Specifically, taking $s = 1/2$, $J_{y} = J_{x} = w_{z} = w_{y} = 0$, $J_{z} = -1$, $w_{x} = h$ we consider (for conciseness we have scaled the terms in Eq. (\ref{Eq:SMGeneralHamiltonian}) -- none of the Physics is changed) the Hamiltonian for the total energy:
\begin{equation}
    \hat{H}_{\rm TFI}(\mathcal{G}) = -\frac{L}{N_{E}}\sum_{(v,v') \in E}\hat{\sigma}_{v}^{z}\hat{\sigma}_{v'}^{z} + h\sum_{v \in V}\hat{\sigma}_{v}^{x}.
    \label{Eq:SMTFIHamiltonian}
\end{equation}
\par In Supplementary Fig. \ref{Fig:SMF3} we use Matrix Product State simulations to compare results from the ground state of this Hamiltonian on the maximally irregular graph to results for the ground state properties on the ER graph. Our finite-size numerics for the ER graph demonstrate convergence to exact results for $L = \infty$, which correspond to the well-known solution to the TFI model on the complete graph \cite{CompleteTFI}. The second order phase transition which is manifest on the ER graph is thus directly underpinned by the system behaving like a single collective spin -- the randomness and inhomogeneity inherent in the actual adjacency matrix plays no role in the Physics of the system. This is further evidenced by the  diminishing of the Shannon entropy of the two-point correlations with increased system size. 
\par Whilst a second order transition between ferromagnetic and paramagnetic phases is also manifest in the ground-state on the maximally irregular graph the functional behaviour of observables in the system is distinct from the ER graph. Moreover, the Shannon entropy of the two point correlations is not diminishing with system size, significant along all spin axes, maximised near the critical point and occurs in unison with a non-zero entanglement entropy in the system. These results serve as further numerical evidence of Theorem \ref{Theorem:SMTH1} and reinforce our conclusions from the main text, emphasizing that they are not specific to the XXZ limit of $\hat{H}(\mathcal{G})$.
\par We note that we added a small term of the form $\frac{1}{200L} \sum_{v \in V}\hat{\sigma}^{z}$ to Eq. (\ref{Eq:SMTFIHamiltonian}) when performing the DMRG simulations to break the $Z_{2}$ symmetry and ensure the longitudinal magnetisation and entanglement entropy are well-defined --- all other physical observables are affected negligibly by this term. We also assume such a term, which is negligible other than breaking the $Z_{2}$ symmetry and forcing $\langle \hat{S}^{z} \rangle \geq 0$, has been added to the Hamiltonian in the ER case for $L = \infty$ in order to calculate the longitudinal magnetisation density.

\begin{figure}[!t]
    \centering
    \includegraphics[width =\textwidth]{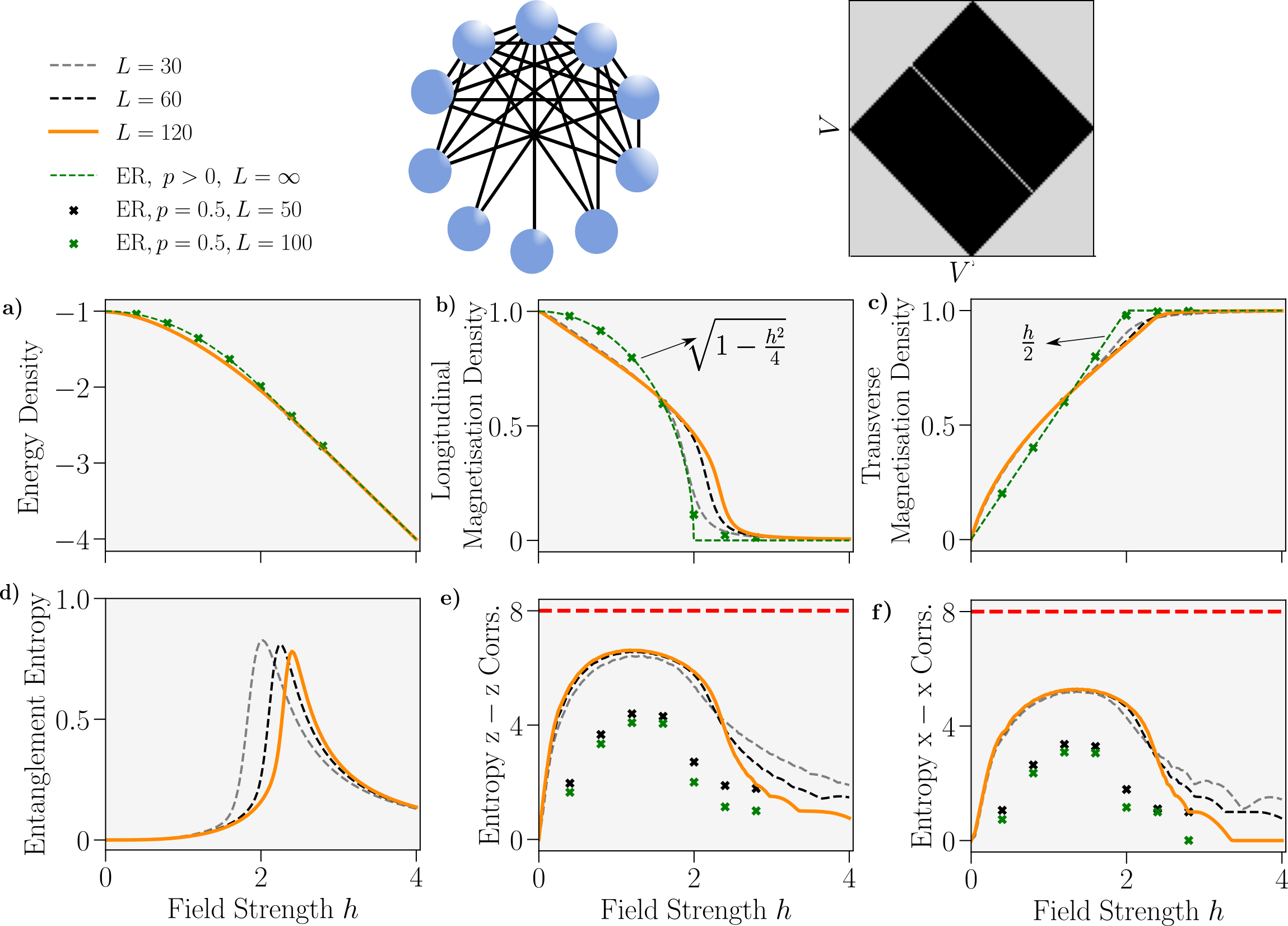}
    \captionsetup{width=\textwidth}
    \caption{Properties of the ground-state of the spin-$\frac{1}{2}$ TFI Hamiltonian -- see Eq. (\ref{Eq:SMTFIHamiltonian}) -- on the `Maximally Irregular Graph' where each pair of vertices, other than two, have a different degree. Graph schematic alongside adjacency matrix for $L= 100$ is provided. System sizes are coded by colour and a full legend is provided top left.
    \textbf{a-f)} Ground-state energy density, longitudinal magnetisation density $\langle \hat{S}^{z} \rangle / L$, transverse magnetisation density $\langle \hat{S}^{x} \rangle / L$, von-Neumann entanglement entropy between sites $V = 1 ... L/2$ and $V = L/2 + 1 ... L$ and Shannon-Entropy of the $z-z$ and $x-x$ correlations (see Methods) versus transverse field strength $h$. We use $n = 256$ bins to calculate the Shannon Entropy. A dotted red line has been added to plots e) and f) to indicate the maximum possible value for the Shannon Entropy with the numbers of bins used. Bond dimensions of $\chi = 250, 500$ and $500$ have been used for system sizes $L = 30, 60$ and $120$ respectively. For comparison, results have been added for the ER graph: the green dashed line represents the exact expression for ground-state observables of $\hat{H}_{\rm TFI}(\mathcal{G}_{\rm ER}(p))$ for any finite non-zero $p$ in the thermodynamic limit and the black and green markers represent DMRG calculations for the ER graph with $p = 0.5$ and $L = 50$ and $L = 100$ respectively. The finite-size ER results were averaged over $n = 10$ draws of the ER graph from its ensemble. A bond dimension of $\chi = 500$ was used for all finite-size calculations --- which we found sufficient to reach our desired accuracy.}
    \label{Fig:SMF3}
\end{figure}

\clearpage

\makeatletter
\renewcommand\@bibitem[1]{\item\if@filesw \immediate\write\@auxout
    {\string\bibcite{#1}{S\the\value{\@listctr}}}\fi\ignorespaces}
\def\@biblabel#1{[S#1]}
\makeatother

\renewcommand\refname{References for Supplementary Information}

\end{document}